\documentstyle[aps,pre,eqsecnum,multicol,epsf]{revtex}
\begin{document}
\renewcommand{\thesection}{\arabic{section}}
\title{Statistical Description of Acoustic Turbulence}
\author {V.S L'vov\cite{lvov},  Yu. L'vov\cite{yuri}, A.C. Newell
\cite{newell} and V. Zakharov\cite{zakharov}}
\address{
\cite{lvov}~~Departments of Chemical Physics, The Weizmann
Institute of Science, Rehovot 76100, Israel,\\ 
\cite{yuri,newell,zakharov}~~
Department of Mathematics,   The University of Arizona, 
        Tucson, Az, 85721, USA   \\
\cite{zakharov}~~Landau Institute for   Theoretical 
Physics, Russian Acad.  of Sci., 117940, GSP-1, Moscow, Kosigina 2, 
Russia} 
\maketitle
\begin{abstract}
  We develop expressions for the nonlinear wave damping and frequency
  correction of a field of random, spatially homogeneous, acoustic
  waves.  The implications for the nature of the equilibrium spectral
  energy distribution are discussed.
\end{abstract} 
\pacs{PACS numbers }  
\begin{multicols}{2}
\section{Introduction and General Discussion}
Weak or wave turbulence, which describes the behavior of a spatially
homogeneous field of weakly interacting, random dispersive waves, has
led to spectacular success in our understanding of spectral energy
transfer processes in plasmas, oceans and planetary atmospheres
\cite{ZLF}.  Furthermore, the subject provides a useful paradigm for
helping one think about some of the challenges of fully developed
turbulence.  First and foremost, the equation for the long time
behavior of the spectral cumulants (which are in one to one
correspondence with the spectral moments) are {\it closed} without
making a priori and unjustifiable assumptions on the statistics of the
processes (such as the quasi-Gaussian  approximation).  Second, the
kinetic equation, which describes the spectral energy transfer via
$n$-wave resonant processes, admits classes of {\it exact} equilibrium
solutions that can be identified as pure Kolmogorov spectra, namely
equilibria for which there is a constant spectral flux of one of the
conserved densities (e.g. energy, number density) of the physical
process under consideration.  By contrast, the thermodynamic
equilibria, which have very little relevance in any turbulence theory
that must account for a sink at small scales, have zero flux. Third,
the theory allows one to glimpse the origin of the intermitency and
the breakdown of the conditions under which one can expect relaxation
to one of the finite flux Kolmogorov equilibria.

The basic ideas for writing down the kinetic equation to describe how
weakly interacting waves share their energies go back to Peierls but
the modern theories have their origin in the works of Hasselman \cite
{Hass}, Benney and Saffmann \cite{Ben}, Kadomtsev \cite{K}, Zakharov
\cite{ZLF}, Benney and Newell \cite{Newell,Newell1}.  A
particularly important event in this history was the discovery of the
pure Kolmogorov solution by Zakharov \cite{Zkolm}.  Usually, the
thermodynamic equilibrium solutions can be seen from the kinetic
equation by inspection.  On the other hand, the solutions,
corresponding to pure Kolmogorov spectra are much more subtle and only
emerge after one has exploited scaling symmetries of the dispersion
relation and the coupling coefficient via what is now called the
Zakharov transformation\cite{ZLF}.

But success to this point, namely the natural closure, 
depended  crucially on the fact the waves were 
dispersive. This means that the group velocity is neither  constant  
in amplitude nor direction, or alternatively 
stated, the dispersion tensor
\begin{equation}
D_{\alpha\beta}=\left( \frac{\partial ^2 \omega}
{\partial k_\alpha \partial k_\beta}\right)
,\ \ \ \ \ 0<\alpha, \beta \leq \alpha\ \ \label{N1} \end{equation}  
has full rank.  Here $d$ is the system dimension, Greek letters 
(here $\alpha$ and $\beta$)   
denote tensor indices  varying from 1 to space dimension $d$, 
and 
\begin{equation} 
\omega=\omega({\bf k})
\label{N2}
\end{equation}
is the linear dispersion relation. The reason 
for closure is slaving. 
In a field of weakly interacting random dispersive waves,
the first step to slaving is achieved by the linear characteristics of
the wave trains.  Systems, which initially are strongly correlated,
are decorrelated because different waves carry statistically
independent information from long distances at different speeds and
directions.  Cumulants of order $N>2$ tend to zero
on the fast time scale $\omega^{-1}_{\rm I}$ ($\omega_{\rm I}$ is a
typical frequency at which the energy is injected). The system
approaches a state of exact joint Gaussian statistics.  The energy at
each wavevector remains constant and there is no transfer. But the
systems of interest to us are nonlinear and therefore, although the
cumulants undergo an initial decay, they are regenerated by the
nonlinear terms. In particular, the cumulant of the order $N$ is
regenerated  both by cumulants of higher order and by  
and products of lower order cumulants.  The
second important reason for closure is the following. The important
terms in the regeneration of the $N^{\rm th}$ order cumulant are not
the terms containing cumulants of order higher then $N$ but rather
those terms which are products of lower order cumulants.  Important
means that even though the nonlinear coupling is weak, the effects of
these terms persist over long times 
because of resonant interaction.  Furthermore, the
regeneration process takes place on a much longer time scale than does
the initial decorrelation process due to wave dispersion.  On this
long time scale, namely the time taken for triad or quartet (or, as in
some rare cases, quintic) resonances to produce order one effects, the
system of equations for the cumulant hierarchy becomes closed.  If $
\epsilon $ is a typical dimensionless wave amplitude (for acoustic
waves it is $\delta\rho /\rho_0$, the ratio of average fluctuation
density amplitude to the ambient value) then this time (measured in
units of the timescale $\omega_{\rm I}^{-1}$) is $ \epsilon ^{-2}$ for
triad resonances and $ \epsilon ^{-4}$ for quartet resonances,
although there is an additional frequency correction in the latter
case that comes in on the $ \epsilon ^{-2}$ time scale.

Mathematically, these results are obtained by perturbation theory, in
which the terms leading to long time cumulative effects can be
identified, tabulated and summed. The method closely parallels that of
the Dyson-Wyld diagrammatic approach which will be discussed in
Section 4. A key part of the analysis is the asymptotic 
($\lim_{t\to \infty}$) evaluation of
certain integrals such as 
\begin{equation} 
\int f({\bf k}_r)\Delta\left[\sum_{r=1}^Ns_r\omega({\bf k}_r)\right] 
\delta\left(\sum_{r=1}^N{\bf k}_r\right )\Pi d {\bf k}_r\,,
 \label{N3}
\end{equation}
where 
\begin{equation} 
{\Delta(h)= \int_0^t dt \exp (i h t) = \frac{\exp (i h t)-1}{i h}}\,,
\label{N4} 
\end{equation} 
and $\delta(x)$  is the Dirac delta function.
 The function $\Delta(h)$ contains the fast (oscillations of the order
of $\omega_I^{-1}$) time $t$, whereas the other functions in the
integrand, here denoted by $f({\bf k}_r)$, only change over much
longer times.  The exponent of $\Delta(h)$ is $\sum_1^{N} s_r
\omega({\bf k}_r)$ where $\omega({\bf k}_r)$ is the linear dispersion
relation and $s_r$ (often $s_r=\pm 1$) denotes its multiplicity.  For
example, in acoustic waves, a wavevector $\bf k$ has two frequencies
corresponding to waves running parallel and anti-parallel to ${\bf k}$.
 The maximum contribution to integrals such as (\ref{N3}) in
the limit of large time $t$ occurs on the so called resonant manifold
$M$, where
\begin{equation}
\sum_{r=1}^{N} {\bf k}_r = 0 \,,
\qquad
h=   \sum_{r=1}^{N}s_r\omega( {\bf k}_r) =0\,,
\label{N6}
\end{equation}
for some choices of the sequence $s_r$. However, the precise form of
the asymptotic limit also depends on whether the zeros of $h$ on $M$
are simple or of higher order.  For the case of (fully) dispersive
waves, such as gravity waves on deep water, Rossby waves, waves of
diffraction on optical beams, the zero of $h$ is simple and (for
sufficiently smooth $f$) one has
\begin{eqnarray}
&& \lim_{t\rightarrow \infty} \int_{-\infty}^{\infty}f(h)\frac
{\exp(i h t) -1 } {ih} d h 
 \nonumber \\
&=& \pi\, {\rm sgn} (t) f(0) + i P \int_{-\infty}^{\infty} 
\frac{f(h)}{h}d  h
 \label{N7}
 \end{eqnarray}
or schematically,
\begin{equation}
\Delta(h)\propto \pi \,{\rm sgn}(t) \delta(h)
 + i P(\frac{1}{h}), \label{N8} \end{equation}
where $P$ denotes Cauchy principal value. In these cases, the
integrand in the kinetic equation, the equation describing the
resonant transfer of spectral density, contains products of energy
densities and the Dirac delta-functions $\delta(\sum_1^N{s_r\omega({\bf
k}_r)})$ and $\delta (\sum_1^N {\bf k}_r)$ clearly indicating that
spectral energy transfer takes place on the resonant manifold $M$. The
asymptotic equations for the change of the higher order cumulants can
be interpreted as a complex frequency modification whose real part
describes the expected nonlinear shift in frequency and whose
imaginary part describes a broadening of the resonant manifold along
its normal directions.

But acoustic waves are not fully dispersive. The linear dispersion
relation,
\begin{equation}
{\omega({\bf k})=c|k|=c\sqrt{{k_{\parallel }^2+k_{\perp}^2}}\,,
 \quad {{\bf k}} 
= (k_{\parallel},k_{\perp})}
\label{N9} 
\end{equation}
where $c$ is the sound speed, leads to a dispersion tensor which has
rank $(d-1)$.  As we will see, this changes the asymptotic.
Furthermore, three wave resonances occur between wavevectors which are
purely collinear. Therefore, since the kinetic equation (KE) only
considers wave interaction on the resonant manifold, there is no way
of redistributing energy out of a given direction.  At best, the KE
will only describe spectral energy transfer along rays in wavevector
space.  Moreover, depending on dimension $d$, the long time behavior
of the integrals (\ref{N4}) differ greatly.  For a given vector $\bf
k$, the locus of the resonant partners ${\bf k}_1$ and ${\bf k} - {\bf
k}_1$ in a resonant triad is given by the surface in ${\bf k}_1 $
space defined by
\begin{equation}
h({\bf k}_1)= s_1 k_1 + s_2 |k - k_1| - s |k| =0\ .
\label{N11}
\end{equation}
Here $s,s_1,s_2=\pm 1$. For $d=1$ and the appropriate choices of the
wave directions $s_1,\ s_2$ and $s$, this manifold is {\it all} ${\bf
k}_1$.  Therefore the fast oscillations in the integral are of no
consequence and do not cause any decorrelation to occur.  All waves
moving in the same directions travel with the same speed. Initial
correlations are completely preserved.  Moreover, we know that for one
dimensional compressible flow, nonlinear terms, no matter how weak
initially, eventually lead to finite time multivalued solutions.
Assuming the usual viscous regularization, multivalued solutions are
replaced by shocks, namely almost discontinuous solutions where
discontinuities are resolved across very thin viscous layers.  One
would naturally expect an energy spectrum $E_1(k)$ which reflects this
fact, namely
\begin{equation}
E_1(k)\propto 1/k^2\ .
\label{N12}
\end{equation} In two 
dimensions, one has dispersion (diffraction) in one direction.
 Indeed, for $d>1$, while
\begin{equation}
\nabla_{{\bf k}_1} h = 0 
\label{N13}
\end{equation}
on the manifold $M$, the Hessian of $h({\bf k}_1)$ is not identically
 zero. In two dimensions, the integral (\ref{N4}) behaves as \FL
\begin{equation}
\int f(x) \frac{\exp (i x^2 t) -1 }{ i x^2} d x \propto 2 t \int f(x)
\exp (i x^2 t) d x,
\label{N14}
\end{equation} 
which grows as $t^{1/2}$ as $t\rightarrow \infty$. In three
dimensions, the growth is much weaker.  Since that is the case we will
look at in detail, we give the exact result. Let
\begin{eqnarray}
{\bf k} &=& (K>0,0,0)\,,\ {\bf k}_1 
= (K_x,K_y,K_z)\,,
\nonumber \\
{\bf k}_2 &=& (K - K_x, -K_y, -K_z)\ .\nonumber \end{eqnarray}
Then, for $s_1=s_2=s,$ 
\begin{eqnarray}
h &=& c(s_1 |{\bf k}_1|+s_2|{\bf k} - {\bf k}_1|- s K )
\label{N15}\\
& =& 
\frac{s K c }{ 2 K_x (K - K_x)}(K_y^2+K_z^2)+O(K_y^3, K_y^2K_z, 
\dots)
\nonumber
\end{eqnarray}
 near the resonant value $(K,0,0)$. The integral 
$$
\int_{-\infty}^{\infty} f(K_x, K_y, K_z;s_1, s_2) \frac{e^{i h t} - 
1}{ i h } d K_x d K_y d K_z 
$$
tends to
\begin{eqnarray}
&& {\alpha} \int_{-\infty}^{0}f(K_x,0,0;-s,s)(-K_x)(K-K_x)d K_x 
\nonumber \\
&+& {\alpha} \int_{0}^{K} f(K_x,0,0;s,s)K_x(K-K_x)d K_x 
\nonumber \\
&+ & {\alpha} \int_{K}^{\infty}f(K_x,0,0;s,-s)K_x(K_x-K)d K_x 
\label{N16}\\
&-& \frac{ 2 i{\alpha} s }{\pi}\log t \int_{-\infty}^{0} 
f(K_x,0,0,;-s,s)K_x(K-K_x)d K_x 
\nonumber \\
&-& \frac{2 i {\alpha}  s}{\pi}\log t\int_0^K 
f(K_x,0,0;s,s)K_x(K-K_x)d K_x  
\nonumber \\
& - & \frac{2 i {\alpha} s }{\pi}\log t \int_{K}^\infty 
f(K_x,0,0;s,-s)K_x(K-K_x)
d K_x 
\nonumber
 \end{eqnarray} 
in the limit $t\rightarrow\infty$. Here $\alpha=\pi^2/ K c $ and we
have kept only the leading order real and imaginary contributions. The
essential difference with (\ref{N7}) and (\ref{N8}) is the additional
Dirac delta function multiplied by $\log t$ in the imaginary
term. This will not change the kinetic equation for the spectral
energy density. If we write the total energy per unit volume $E$ as
\begin{equation} 
E= 2 \rho_0 c^2  \epsilon ^2 \int e({\bf k}) d {\bf k} 
\label {N17}
\end{equation}
where $\rho_0$ is the ambient density and $ \epsilon $ 
a measure of  amplitude, then 
\begin{eqnarray}
&&\frac{d e ({\bf k})}{ d t } = St(e,\dot{e})\nonumber\\     
&&St(e,\dot{e})=
\frac{ \pi^2 c (\mu+1)^2  
\epsilon ^2 K^4}{4}
\Bigg\{2\int^\infty_0 d  \gamma  \gamma ( \gamma+1)
\nonumber \\
&\times&
\Big[ e( \gamma{\bf k})e(( \gamma+1){\bf k})
+ \gamma e({\bf k}) e(( \gamma+1){\bf k})
\nonumber \\
&-&( \gamma+1)e({\bf k})e( \gamma {\bf k})\Big]
+\int_0^1d k \alpha(1-\alpha)\Big[ e({\alpha} k) e((1-\alpha){\bf k})
\nonumber \\
&-& {\alpha} e({\bf k})e((1-\alpha){\bf k})
-(1-\alpha)e(k)e({\alpha} {\bf k}) \Big] \Bigg\}
\label{N18}
\end{eqnarray}
where $\mu$ is the adiabatic constant [$p=p_0(\rho/\rho_0)^\mu$] and
$|{\bf k}|=K$. In $d$ dimensions a little calculation
show, that the RHS of (\ref{N18}) has the $t$ dependence
$t^{\frac{3-d}{2}}$ so that in general the nonlinear interaction time
$\tau_{NL}$ for the resonant exchange of spectral energy is 
$\epsilon^2 t^{\frac{5-d}{2}}=O(1)$ 
or $\tau_{NL}\propto \epsilon^{-\frac{4}{5-d}}$.
 (Note that for $d\ge 5$, there is no
cumulative effect of this resonance.)

While the extra term in (\ref{N16}) proportional to $i \log t$ plays no
role in the spectral energy transfer, it will, however, appear in the
frequency modification.
Calculating the long time behavior of the higher order 
cumulants leads to a natural re-normalization of the frequency, 
\begin{eqnarray} 
&&\omega({\bf k})= c |{\bf k}| \Bigg[1-2 \pi (\mu+1)^2  \epsilon ^2 
\ln{\frac{1}{ \epsilon ^2}} 
\int_0^\infty\beta^2 e(\beta \hat k) d \beta 
\nonumber \\
&+& O( \epsilon ^2)\Bigg] + 
i \pi ^2 (\mu+1)^2  \epsilon ^2 \Bigg[\int_{|{\bf k}| 
}^{\infty}\beta^2 e(\beta \hat k) d \beta 
\nonumber \\
&+& \frac{1}{|{\bf k}|}\int_0^{| {\bf k}| } 
 \beta^3 e(\beta \hat k) d \beta 
+ |k| \int_0^{|{\bf k}| } \beta e(\beta \hat k) d \beta 
 \Bigg]
\label{N19}
\end{eqnarray} 
where $\hat k = {\bf k}/{K}$. The calculation of the frequency
re-normalization is the new result of this paper.  We present two
derivations of this result, in the framework of the above analysis and
making use of a diagrammatic perturbation approach.

The equation (\ref{N18}) is nothing but a ``regular" kinetic equation
for the three-wave interactions, written in a dispersionless limit
$\omega = c |{\bf k}|$. In this case three wave resonant conditions
\begin{eqnarray}
\pm \omega({\bf k})=\pm\omega({\bf k_1})\pm\omega({\bf k_2})\ \ \ \ \ 
 {\bf k}={\bf k_1}+{\bf k_2}\label{ZAH1rez}\end{eqnarray} can be
satisfied if and only if all three vectors $ {\bf k}, {\bf k_1}, {\bf
k_2}$ are parallel, as a result, the integration over ${\bf k_1},{\bf
k_2}$ is along line parallel to $\bf k$. It is unclear a priori that
the three wave kinetic equation can be used in the dispersionless
case; is certainly less plausible in the two dimensional case where
the formal implementation of the kinetic equation leads to stronger
divergences.

The derivation presented above is taken from the article of Newell and
Aucoin \cite{NeAu}, who made the first serious attempt of an
analytical description of the dispersionless acoustic turbulence.

Newell and Aucoin \cite{NeAu} also argued that a natural asymptotic
closure also obtains in two dimensions because of the relative higher
asymptotic growth rates of terms in the kinetic equation involving
only the spectral energy, but this is still a point of dispute, is not
yet resolved and will not be addressed further here.

Independently the kinetic equation (\ref{N18}) was applied to acoustic
turbulence by Zakharov and Sagdeev \cite{ZS} who used it just as a
plausible hypothesis.  However, Zakharov and Sagdeev also suggested an
explicit expression for the spectrum of acoustic turbulence
\begin{equation} e(k)\propto k^{-3/2} \label{Z2spec}\end{equation}
which is just a Kolmogorov-type spectrum, first obtained by Kolmogorov
from dimensional considerations in the context of hydrodynamic
turbulence.  Here, however, the (\ref{Z2spec}) is an exact solution of
the equation
\begin{equation} St(e,\dot{e})=0.\label{Z3KE}\end{equation}
The proof of this fact can be found in the \cite{ZLF}. One should also
mention, that the quantum kinetic equation applied to a description of
a system of weakly interacting dispersionless phonons were done as
long ago as in 1937 by Landau and Rumer (\cite{LR}).

Kadomtsev and Petviashvili \cite{KP} criticized this result on the
grounds that the kinetic equation is in the dispersionless case can
hardly be justified because of the special nature of the linear
dispersion relation.  They suggested that acoustic turbulence in two
and three dimensions was much more likely to have parallels with its
analogue in one dimension. We have already mentioned in that case that
the usual statistical description is inadequate both because there is
no decorrelation dynamics and because shocks form no matter how weak
the nonlinearity initially is. The equilibrium statistics relevant in
that case is much more likely to be a random distribution of
discontinuities in the density and velocity fields which lead to an
energy distribution of (\ref{N12}). Further, Kadomtsev and
Petviashvili argued that even in two and three dimensions one would
expect the same result, namely
\begin{equation} k^{d-1}e({\bf k})\propto k^{-2}\label{N22}
\end{equation} a random distribution of statistically independent
shocks propagating in all directions.

But wave packets traveling in almost parallel directions are not
independent.  Consider a solid angle containing
$N=(k_{\parallel}/k_{\perp})^{d-1}$ wavepackets with wavevectors
$(k_{\parallel},k_\perp)$ where $k_{\parallel}= l^{-1}$ is a typical
length scale of the fluctuating field in the direction of the
propagation, and $k_\perp \ll k_\parallel$.  The shock time $\tau_{\rm
  sh}$ for a single wave packet would be $l\sqrt{{\rho}N/E}
\propto(l/c\epsilon)N^{(1/2)}$, where $E$ is the total energy in the
field. The dispersion (diffraction) time $\tau_{\rm disp}$, namely the
time over which several different packets have time to interact
linearly, is of the order of $k_{\parallel}/(c k^2_{\perp})\propto l
N^{2/(d-1)}/c$.  As we have already observed, the nonlinear resonance
interaction time $\tau_{NL}$ for spectral energy transfer is
$(l/c)\epsilon^{-({4}/({5-d}))}$. The ration is
$\tau_{disp}:\tau_{sh}:\tau_{NL}=N^{{2}/({d-1})}:N^{{1}/{2}}
\epsilon^{-1}:\epsilon^{-{4}/({5-d})}$. In the limits
$N\to\infty,\,\,\, \epsilon\to 0$, the shock time is sandwiched
between the linear dispersion time and nonlinear interaction time and,
if we choose $N(\epsilon)$ by equating the first two, all three are
the same. Moreover, the phase mixing which occurs due to the crossing
of acoustic wave beams, occurs on a shorter time scale, a fact that
suggests that the resonant exchange of energy is the more important
process. But even then, several very important questions remain.
\begin{enumerate}
\item To what distribution 
does the energy along a given wavevector ray 
relax? 
\item How does energy become shared between neighboring rays?
\item Does energy tend to diffuse away from the ray with maximum
energy or can it focus onto that ray? In the latter case, one might
argue that shock formation may again become the relevant process
especially if the energy should condense on rays with very different
directions.
\end{enumerate}

The aim of this paper is to  take a very modest first step in the 
direction of answering these questions. In particular, we present a 
curious result. The fact that there is a strong 
($ \epsilon ^2 \ln 1/ \epsilon ^2$) 
correction to the frequency leads us to ask if that terms could
provide the dispersion required to allow the usual triad resonance
process carry energy between neighboring rays. At first sight, it
would appear that that is indeed the case, that the modified nonlinear
dispersion law is
\begin{equation}
\omega({\bf k})= c({\bf k} ) ( 1 +  \epsilon ^2 \ln{\frac{1}{ \epsilon ^2}
\Omega(k)})
\label{displaw}
\end{equation}
where $\Omega$ is proportional to $|{\bf k}|$. But a surprising and
nontrivial cancellation occurs which means that the first corrections
to the wave speed still keeps the system non dispersive in the
propagation direction.

While this fact is the principal new result of this paper, our
approach lays the foundation for a systematic evaluation of the
contribution to energy exchange that occurs at higher order. Indeed,
we expect that some of the terms found by Benney and Newell
\cite{Newell} involving gradients across resonant manifolds which, in
the fully dispersive case, are not relevant because the resonant three
wave interaction gives rise to an isotropic distribution, may be more
important in this context.
\vskip .5cm

{\bf The paper is written as follows.}  In the next Section, we derive
the equation of motion for acoustic waves of small but finite
amplitude.  A second approach discussed in Subsect. 2B starts from the
Hamiltonian formulation of the Euler equations and again makes use of the
small  amplitude parameter of the problem to simplify 
the interaction Hamiltonian. As we will see in Subsect. 2C both
approaches are equivalent and which approach to use is the question of
taste.

Next, in Section 3 we write down the hierarchy of equations for the
spectral cumulants and solve them perturbatively. Certain resonances
manifest themselves as algebraic and logarithmic time growth in the
formal perturbation expansions and mean that these expansions are not
uniformly asymptotic in time. The kinetic equation, describing the
long time behavior of the zeroth order spectral energy, and the
equations describing the long time behavior of the zeroth order higher
cumulants are simply conditions that effectively sum the effect of the
unbounded growth terms. Under this renormalization, the perturbation
series becomes asymptotically uniform. By asymptotically uniform, we
mean that the asymptotic expansion for each of the cumulants remains
an asymptotic expansion over long times. All unbounded growths are
removed. While this procedure in principle requires one to identify
and calculate unbounded terms to all orders, in practice one gains a
very good approximation by demanding uniform asymptotic behavior only
to that order in the coupling coefficient where the unboundedness
first appears.

In other words this means that if one finds that if the first two
terms of the asymptotic expansion are $1+ \epsilon ^2 t \psi_1+ ...$,
then the effective removal of $\psi_1$ will remove all terms which are
powers of $( \epsilon ^2 t)$ in the full expansion.  Likewise, it also
assumes that there appear no worse secular terms at a higher order,
such as for example $ \epsilon ^4 t^3 \psi_2$.  To achieve uniformity,
one requires an intimate knowledge of how unbounded growth appears.
This sort of perturbative analysis was first done in the thirties by
Dyson. A technical innovation was to use graph notations, called {\it
diagrams}, for representing lengthy analytical expressions for high
order terms in the perturbation series. It often happens that one can
find the principal subsequence of terms just by looking on the
topological structure of corresponding diagrams. This method of
treating perturbation approaches is called {\it the diagrammatic
technique}.

The first variant of diagrammatic technique for non-equilibrium
processes was suggested by Wyld\cite{W961} in the context of the
Naiver Stokes equation for an incompressible fluid.  This technique was
later generalized by Martin, Siggia and Rose \cite{MSR3}, who
demonstrated that it may be used to investigate the fluctuation
effects in the low-frequency dynamics of any condensed matter system.
In fact this technique is also a classical limit of the Keldysh
diagrammatic technique \cite{K964} which is applicable to any physical
system described by interacting Fermi and Bose fields.  Zakharov and
L'vov \cite{ZL75} extended the Wyld technique to the statistical
description of Hamiltonian nonlinear-wave fields, including
hydrodynamic turbulence in the Clebsch variables \cite{L991}. In
section 4, we will use this particular method for treating acoustic
turbulence.

Note that in such a formulation, unbounded growths appear as
divergences (or almost divergences) due to the presence of zero
denominators caused by resonances, the very same resonances, in fact,
that give rise to unbounded growth in our more straightforward
perturbation approach.  Moreover, diagrammatic techniques are
schematic methods for identifying all problem terms and for adding
them up.  If one uses the diagram technique only to the first order at
which the first divergences appear, this is called the one-loop
approximation and is equivalent to identifying the first long time
nonlinear effects.  This is exactly analogous to what we will do in
our first approach in this paper although we will also display the
diagram technique.  The one loop approximation will give the same long
time behavior of the system for times of $\tau_{NL}$ defined earlier.
In Appendix C we analyze two loop diagrams and show that some of them
gives the same order contribution to $\gamma_k$ as two loop diagrams.
Nevertheless one may believe, that even one-loop approximation gives
qualitatively correct description of the dynamics of the system.

The last Section 5 is devoted to some concluding remarks and the
identification of the remaining challenges.  We now begin with
deriving the basic equations of motion for weak acoustic
turbulence. 
\section{Basic equation of motion for weak acoustic turbulence} 
\subsection{Straightforward Approach}
Consider the Euler equations for a compressible fluid:
\begin{eqnarray}
&& \partial \rho/\partial 
t+{\bbox\nabla}\cdot\,(\rho{\bf v})=0\,,
\nonumber \\
&&\partial{\bf v}/\partial t+({\bf v}\cdot {\bbox{\nabla}}){\bf v}=
        -{\bbox\nabla} p(\rho)/\rho\ .
        \label{B1}
\end{eqnarray}
Here $ v({\bf x},t)$ is the Euler fluid velocity, $\rho({\bf x},t)$
the density, and $p({\bf r},t)$ is the pressure which, in the general
case, is a function of fluid density and specific entropy $s$
[$p=p(\rho,s)$]. In ideal fluids where there is no viscosity and heat
exchange, the entropy per unit volume is carried by the fluid, i.e.
obeys the equation $\partial s/\partial t+({\bf
v}\cdot{\bbox\nabla})s=0$.  A fluid in which the specific entropy is
constant throughout the volume is called barotropic; the pressure in
such fluid is a single-valued function of density $p=p(\rho)$.  In
this case, ${\bbox\nabla} p/\rho$ may be expressed via the gradient of
specific enthalpy of unit mass $w=E+p V$ and $d w=V d p=d
p/\rho$. Thus, ${\bbox\nabla} p/\rho={\bbox\nabla} w$.

Writing the fluid density $\rho({\bf x},t)$ as $\rho_0(1+\eta({\bf
x},t))$, the velocity field as $v({\bf x},t)$, the pressure field as
$p=p_0(1+\eta)^\mu$ and the enthalpy as
$$w=\int \frac{d p}{\rho} = \frac{c_0^2}{\mu-1}(1+(\mu-1)\eta + 
\frac{(\mu-1)(\mu-2)}{2}\eta^2 + \dots )$$ 
one can write (\ref{B1}) to third order in amplitude
in the following form 
\begin{eqnarray}
\frac{\partial \eta }{\partial t} &+& \frac{\partial v_i}{\partial 
x_i} = - \frac{\partial}{\partial x_i} \eta v_i, 
\label{n1} \\
\frac{\partial v_j}{\partial t} &+& c^2 \frac{\partial \eta}{\partial x} 
= - v_m \frac{\partial v_i}{\partial x_m} - \frac{c^2(\mu-2)}{2}
\frac{\partial}{\partial x_j}\eta^2 
\nonumber \\
&-&\frac{c^2(\mu-2)(\mu-3)}{6} \frac{\partial }{ \partial x_j}\eta^3.
\label{n2}
\end{eqnarray}
Let us introduce new variables as
\begin{eqnarray} 
\eta({\bf x},t) &=& \int \sum_s \epsilon a^s({\bf k},t)e^{i {\bf k}
 {\bf x} + i 
s \omega({\bf k} ) t } d {\bf k}\, ,
\label{n3}\\
v_j({\bf x},t) &=& \int \sum_s \frac{- c^2 k_j}{s \omega({\bf k} ) } 
 \epsilon  a^s({\bf 
k},t) e^{i {\bf k} {\bf x} + i s \omega({\bf k} ) t } d {\bf k}\,,   
\label{n4}
\end{eqnarray}
where  $0< \epsilon \ll 1$, $\omega(\vec  k ) = c |{\bf k} |$ and $\sum_s$
connotes summation over $s=\pm 1$.  From 
(\ref{n1}) and (\ref{n2}), 
\begin{eqnarray}
&&\!\!\!\!\!
\frac{\partial a^s({\bf k}, t)}{\partial t} = \epsilon  \sum_{s_p , s_q} 
\int d {\bf k}_p d {\bf k}_q 
L^{s,s_p,s_q}_{{\bf k},{\bf k}_p,{\bf k}_q} a^{s_p} ({\bf k}_p, t) 
a^{s_q}({\bf k}_q,t) 
\nonumber \\
&\times&\delta({\bf k}_p+{\bf k}_q - {\bf k} ) 
\exp\{i[ s_p\omega({\bf k}_p)+s_q\omega({\bf k}_q)- s \omega({\bf k} )]t\}
\nonumber \\
&+&  \epsilon ^2 \sum _{s_p ,s_q, s_r} \int  d {\bf k}_p
  d {\bf k}_q d {\bf k}_r 
L^{s,s_p,s_q,s_r}_{{\bf k},{\bf k}_p,{\bf k}_q, {\bf k}_r} a^{s_p}
({\bf k}_p,t)  
\nonumber \\ &\times& 
a^{s_q}({\bf k}_q,t)a^{s_r}({\bf k}_r,t) 
\delta({\bf k}_p + {\bf k}_q + {\bf k}_r - {\bf k})
\nonumber \\ &\times& 
\exp\{i[s_p\omega({\bf k}_p)+s_q\omega({\bf k}_q)+s_r\omega({\bf k}_r)- s 
\omega({\bf k})]t\}\label{n5}
\end{eqnarray}                         
where the summation is done over all signs of $s_p,\ s_q,\ s_r$ and we
used the shorthand notation $ \omega_p=\omega({\bf k}_p)$. The coupling
coefficients are,
\begin{eqnarray}
L^{s,s_p,s_q}_{{\bf k},{\bf k}_p,{\bf k}_q} 
&=& \frac{i c^2}{4} \Big(\frac{{\bf k} {\bf 
k}_p}{s_p\omega_p}+\frac{{\bf k} {\bf k}_q}{s_q\omega_q}
+ \frac{s\omega}{s_p\omega_p s_q \omega_q}{\bf k}_p {\bf k}_q \Big) 
\nonumber \\
&& +\frac{i}{4}(\mu-2)s\omega 
\label{n6}\\ 
L^{s,s_p,s_q,s_r}_{{\bf k},{\bf k}_p,{\bf k}_q,{\bf k}_r} &=& 
\frac{i\omega}{12}(\mu-2)(\mu-3) \ .
\label{n7}
\end{eqnarray}
These coefficients have the following important properties:  
\begin{itemize} 
\item $L^{s,s_p,s_q}_{{\bf k},{\bf k}_p,{\bf k}_q} $ 
is symmetric under the interchange of 
$p$ and $q$ 
\item 
$L^{s_p,s,-s_q}_{{\bf k}_p,{\bf k},-{\bf k}_q}= (s_p \omega_p /s\omega)
L^{s,s_p,s_q}_{{\bf k},{\bf k}_p,{\bf k}_q}$ 
\item On the resonant manifold $M$, given by 
\begin{eqnarray} \frac{1}{c}h &=& s_p |{\bf k}_p| + s_q|{\bf k}
 - {\bf k}_p|- s |{\bf 
k}|=0\,, 
\label{n10}\\
L^{s,s_p,s_q}_{{\bf k},{\bf k}_p,{\bf k}_q} &=& \frac{i c s}{4}(\mu+1)K\,,
\label{n11}
\end{eqnarray}
\end{itemize} 
where $|{\bf k}| = K$. Note that if ${\bf k} = (K,0,0)$,
the resonant manifold is not of codimension 
one but degenerates to $K_y=K_z=0$, where ${\bf k}_p = (K_x,K_y,K_z)$, 
 ${\bf k}_q=(K-K_x,-K_y,-K_z)$.  There are three cases. 
\begin{enumerate} 
\item For $K_x<0<K$, $|{\bf k}_p|= - K_x$, $|{\bf k}_q|= K + K_x$,  
$s_p=-s$, $s_q= s$. 
\item For $0<K_x<K$,  $ |{\bf k}_p|=K_x$, $|{\bf k}_q|=K-K_x$ \ $ 
s_p=s_q=s$. 
\item For $0<K<K_x$, $|{\bf k}_p|=K_x$, $ |{\bf k}_q|=K_x-K$, 
$s_p=s,s_q=-s$. 
\end{enumerate}
\subsection{Hamiltonian Description of Acoustic Turbulence}
\subsubsection{Equations of Motion and Canonical Variables}
Consider again the Euler equations for a compressible fluid
(\ref{B1}).  The enthalpy of a unit mass $w=E+pV$ is equal to the
derivative of internal energy of unit volume $\varepsilon(\rho)=E\rho$
with respect to fluid density: $w={\delta\varepsilon/\delta\rho}\ $.
As a result of direct differentiation with respect to time, it is
readily evident that equations (\ref{B1}) conserve the energy of the
fluid
\begin{equation} 
{\cal H}=\int[\rho v^2/2+ \varepsilon(\rho)]\,d{\bf r}\ .  
\label{B3} 
\end{equation} 
One can show (and see for example\cite{ZLF}) that Eqs. (\ref{B1}) may
be written in the Hamiltonian form:
\begin{eqnarray} 
\partial\rho/\partial t&=&\delta{\cal H}/\delta\Phi\;,\quad
\partial\Phi/\partial t=-\delta{\cal H}/\delta\rho\;,
        \label{B4} \\
\partial\lambda/\partial t&=&\delta{\cal H}/\delta\mu\;,
\quad
\partial\mu/\partial t=-\delta{\cal H}/\delta\lambda\;,
\label{B5}
\end{eqnarray}
if the velocity ${\bf v }({\bf r}, t)$ is presented in terms of two 
pairs of Clebsch variables $(\rho,\Phi)$\  and $(\lambda ,\nu)$\  
as follows,
\begin{equation}
{\bf v}=\lambda{{\bbox\nabla}\mu\over\rho}+{\bbox\nabla}\Phi\ .
\label{B6}
\end{equation}
Here the energy (\ref{B3}) is expressed in terms $(\rho,\Phi)\ \ $ and
$ \ (\lambda ,\nu)$ so that (\ref{B6}) becomes the Hamiltonian of the
system.  As seen from (\ref{B6}), the case with $\lambda=0$ or
$\mu=$const corresponds to potential fluid motions which are defined
by a pair of variables ($\rho,\Phi$) according to equations
(\ref{B4}).  It is convenient to transform in the $\bf
k$-representation from the real canonical variables, $\Phi({\bf
k}),\rho({\bf k})$ to the complex ones $b({\bf k})$ and $\ b^*({\bf
k})$,
\begin{eqnarray}
\Phi({\bf  k})&=&-i\sqrt{(c /2\rho_0k)}
[b({\bf  k})-b^*(-{\bf  k})]\,,
\label{A10} \\
\delta\rho({\bf  k})&=&
\sqrt{(\rho_0 k /2\c)}[b({\bf  k})+b^*(-{\bf  k})]\ .
\label{A11}
\end{eqnarray}
Here $\delta\rho({\bf k})=[\rho({\bf k})-\rho_0({\bf k})]$ is the
 Fourier transform of density deviation from the steady state.
\subsubsection{Hamiltonian of Acoustic Turbulence}
Let us expand the Hamiltonian (\ref{B3})
 (expressed in terms of $b,\ b^*$) in power series:
\begin{equation} 
        {\cal H}= {\cal H}_0 +{\cal H}_{\rm int}\ .
        \label{A6} 
\end{equation}
Here  ${\cal H}_0$ is  quadratic in $b$ and $b^*$, giving the 
Hamiltonian of non-interacting waves:
\begin{equation} 
      {\cal H}_0=\int \, c \,  k \, b ({{\bf k}}) b^*({{\bf k}}) d{\bf k},
        \label{A7} 
\end{equation}
with linear dispersion relation $\omega_0({\bf k}) = c k$. In the
Hamiltonian of interaction ${\cal H}_{\rm int}$ we take into account
only three-wave processes:
\begin{eqnarray}
{\cal H}_{\rm int}&=&
{1\over2}\int ( V({\bf  k},{\bf  k}_1,{\bf  k}_2)
 b^*_1b_2b_3+{\rm c.c.})
\nonumber \\
&&\times\delta({\bf  k}_1 -{\bf  k}_2-
{\bf  k}_3)\,d{\bf  k}_1 d{\bf  k}_2d {\bf  k}_3 \ . 
        \label{A9}
\end{eqnarray}
We neglected here $0\leftrightarrow3$ processes (processes described
by $b_1^*b_2^*b_3^*$ and $b_1b_2b_3$ terms), because they are
nonresonant. It means, that if we take into account
$0\leftrightarrow3$ term, it is not going to change our final results,
thus we can neglect it from the very beginning.  We also neglected
contributions from 4-wave and higher terms, because three-wave
interaction is the dominant one.

The coupling coefficient of the  3-wave interaction a given by \cite{ZLF}
\begin{eqnarray}
&&V({\bf k}_0,{\bf  k}_1,{\bf  k}_2)
\label{A13}\\
&=&\sqrt {c k k_1 k_2\over 4\pi^3\rho_0}
   (3g+\cos\theta_{01}+\cos\theta_{02}+\cos\theta_{12})\, , 
\nonumber
\end{eqnarray}
where $g$ is some dimensionless constant of the order of unity and
$\theta _{ij}$ is the angle between ${\bf k}_i$ and ${\bf k}_j$.
Since we have almost linear dispersion relation, only almost parallel
wavevectors can interact, therefore $\cos \theta_{ij}$ with the high
accuracy can be replaced by $1$ and (\ref{A13}) reduces to
\begin{equation}
V({\bf k}_0,{\bf  k}_1,{\bf  k}_2)
\label{A13double}
=\sqrt {c k k_1 k_2\over 4\pi^3\rho_0}
   3(g+1)\, , 
\nonumber
\end{equation}
\subsubsection{Canonical Equation of Motion}
The  Hamiltonian equations of motion (\ref{B4}) for the complex 
canonical variables $b\ b^*$ have  standard form\cite{ZLF}
\begin{equation}
i{\partial b({\bf  k},t)\over\partial t}
={\delta{\cal H}\over\delta
b^*({\bf  k},t)} \  .
\label{A14}
\end{equation}
For the acoustic Hamiltonian (\ref{A6}-\ref{A9}), this equation takes 
the form
\begin{eqnarray}
&&\Big[ i \frac{\partial}{\partial t}
- c k \Big] b({\bf  k},t) 
 \nonumber \\
&=&\frac{1}{2} 
\int V({\bf  k},{\bf q},{\bf  p})b({{\bf q}})b({{\bf  p}})
 \delta({\bf  k}-{\bf q}-{\bf  p}) 
\frac{d {\bf q} d {\bf  p}}{(2\pi)^3} 
\label{A16}\\
&&+\int V^*({\bf  k},{\bf q},{\bf  p})b({{\bf q}})^*b({{\bf  p}}) 
\delta({\bf  k}+{\bf q}-{\bf  p}) 
\frac{d^3 {\bf q} d^3 {\bf  p}}{(2\pi)^3}\ .  
\nonumber
\end{eqnarray}
It is sometimes convenient to concentrate attention on steady state
turbulence, which is convenient to describe in the $ {\bf k},\omega$
-representation.  After performing a time 
Fourier transform, one has instead
of (\ref{A16}),
\begin{eqnarray}
&&
\Big[ \omega - c k \Big]b ({\bf k},\omega)
  =
\frac{1}{2} \int V({\bf  k},{\bf  k}_1,{\bf  k}_2)
\nonumber \\&\times&
b_1 b_2 \delta({\bf  k}-{\bf  k}_1-{\bf  k}_2)
 \delta(\omega-\omega_1-\omega_2)  \frac{d {\bf  k}_1 d
 \omega_1 d{\bf  k}_2 d\omega _2}{(2\pi)^4}
\nonumber \\
&+&  
\int V^*({\bf  k},{\bf  k}_1,{\bf  k}_2)b_1 ^*b_2 
\delta({\bf  k}+{\bf  k}_1-{\bf  k}_2) 
\delta (\omega +\omega_1-\omega_2) 
\nonumber \\&\times& \frac{d {\bf  k}_1 d\omega _1
 d{\bf  k}_2 d\omega _2}{(2\pi)^4}  \ .
\label{A17}
\end{eqnarray}
Hereafter we will refer to this as the {\sl basic equation of motion
  for the acoustic turbulence normal variables $b_k,\ \ b^*_k$} and use it
for
a statistical description of acoustic turbulence.  
\subsection{ Relations between Wave Amplitudes
 $a^+({\bf k}),\ \  a^-({\bf k})$
with Normal 
Variables of Acoustic Turbulence $b({\bf k}),\  b^*({\bf k})$.}
Comparing Eqs. (\ref{n3}) and (\ref{n4}) we get 
\begin{eqnarray}
\delta\rho({\bf k},t) &=&
\rho_0  \epsilon \Big\{  a^+({\bf k},t)\exp [i\omega({\bf k} ) t ]
\nonumber \\
&+&a^-({\bf k},t)\exp [-i \omega({\bf k} ) t ]\Big\}(2\pi)^{3/2}
\label{eq1}\\
\Phi({\bf k},t) 
&=& \frac{i c^2  \epsilon }{ \omega({\bf k} ) } \Big\{a^+({\bf k},t)
\exp[i\omega({\bf k} ) t]  
\nonumber \\
&-&a^-({\bf k},t)\exp[-i\omega({\bf k} )t] \Big\}
(2\pi)^{3/2}
\end{eqnarray}
Here $\Phi $ is velocity potential: ${\bf v} = \nabla \Phi $.  This
gives
\begin{eqnarray} 
a^{+}({\bf k},t)&=&  \frac{\exp [-i\omega({\bf k}) t] }
{2 \epsilon(2\pi)^{3/2}  }
\left[{\delta\rho({\bf k},t)\over \rho_0} -i \Phi({\bf k},t)
 \frac{\omega({\bf k})}{c^2  }\right]\,,
\nonumber \\
a^{-}({\bf k},t)&=& \frac{\exp [i\omega({\bf k}) t] }{2 
\epsilon (2\pi)^{3/2} }\left[{ \delta\rho({\bf 
k},t)\over \rho_0} + i \Phi({\bf k},t)\frac{\omega_k}{c^2}\right]\ .
\end{eqnarray}
Note, that 
$a^+$ and $a^-$ is dimensionless variables.

Now we can easily express $a^+({\bf k}),\ \ a^({\bf k})$ in terms of
$b({\bf k}),\ \ b^*({\bf k})$ and thereby relate the two alternative
approaches presented in this paper,
\begin{eqnarray} 
a^+({\bf k},t) &=& \frac{1}{ \epsilon }\sqrt{\frac{k}{2 c \rho_0}}
(2\pi)^{-3/2}\exp [- i \omega({\bf k}) t] b^*({-\bf k})\,,
 \label{connection1}\\
 a^-({\bf k},t)&=& \frac{1}{ \epsilon }\sqrt{\frac{k}{2 c \rho_0}}
(2\pi)^{-3/2}\exp [i \omega({\bf k}) t] b({\bf k})\ .
\label{connection2}
\end{eqnarray}
To check, that the two approaches are consistent, we rewrite the
equation of motion (\ref{n5}) for $a_k^s$ neglecting $\epsilon^2$
(four-wave interaction) terms:
\begin{eqnarray}
&&\!\!\!\!\!\!
\frac{\partial a^s({\bf k}, t)}{\partial t}=  
 \epsilon  \sum_{s_p s_q} \int d {\bf k}_p d {\bf k}_q 
L^{s,s_p,s_q}_{{{\bf k}},{{\bf k}}_p,{{\bf k}}_q} a^{s_p}
({\bf k}_p, t) a^{s_q}({\bf k}_q,t)
\nonumber \\&\times&
\delta({\bf k}_p+{\bf k}_q - {\bf k} ) 
\exp \{i[s_p\omega({\bf k}_p)+s_q\omega({\bf k}_q)- s \omega({\bf k} )]t\}
\nonumber \\
\label{n5ctruncated}
\end{eqnarray}
Now we substitute Eqs.  (\ref{connection1}) and (\ref{connection2})
into (\ref{n5ctruncated}) and obtain
\begin{eqnarray}
&&\!\!\!\!\!\!
\left[{\partial \over \partial t} + i \omega({\bf  k})\right]
  b({{\bf  k}},t)= -i 
\int d{\bf  p} d{\bf q}\sqrt{\frac{k p q c }{4\pi^3\rho_0}} 
\nonumber \\&\times&
\Big[ (\mu-2) +\cos{\theta_{{\bf  k},{\bf  p}}}
+\cos{\theta_{{\bf  k},{\bf q}}}
+\cos{\theta_{{\bf  p},{\bf q}}}\Big]  
\label{eqmotioncomp}\\
&\times&
\left[\delta({{\bf  k}+{\bf  p}+{\bf q}})b_{{\bf  p}}^*b_{{\bf q}}^* + 
2 \delta_{{\bf  k}+{\bf  p}-{\bf q}}b_{{\bf  p}}^*b_{{\bf q}} 
        +   \delta({{\bf  k}-{\bf  p}-{\bf q}})b_{{\bf  p}} b_{{\bf q}}
 \right] 
\nonumber
\end{eqnarray}
Now one can see that equation (\ref{eqmotioncomp}) looks exactly as
(\ref{A14}) with Hamiltonian (\ref{A9}) and with coupling coefficient
(\ref{A13}). 
Thus one conclude that the two approaches are equivalent and the choice
between them is the question of taste.

\section{Long-time Analysis of statistical behavior}
The analysis proceeds by first forming the hierarchy of equations for
the spectral cumulants (correlation functions of the wave amplitudes)
defined as follows.  The mean is zero.
\begin{eqnarray}
&&\left<a^s({\bf k})a^{s'}({\bf k}')\right>
=\delta({\bf k}+{\bf k}')q^{s s'}({\bf k}, {\bf k}'),
\label{normal}\\
&&\left<a^s({\bf k})a^{s'}(k')a^{s''}({\bf k}'')\right>
\nonumber \\
&=&\delta({\bf k}+{\bf k}'+{\bf k}'') q^{s s' s''}
({\bf k},{\bf k}',{\bf k}''),
\label{abnormal}\\
&&\left<a^{s}({\bf k})a^{s'}({\bf k}')
a^{s''}({\bf k}'')a^{s'''}({\bf k}''')\right> 
\nonumber \\
&=&\delta({\bf k}+{\bf k}'+{\bf k}''+{\bf k}'''
)q^{s s' s'' s'''}({\bf k},{\bf k}',{\bf k}'',{\bf k}''') 
\nonumber \\
&&+\delta({\bf k}+{\bf k}')\delta({\bf k}''+{\bf k}''')
 q^{s s'} ({\bf k},{\bf k}') 
q^{s''s'''}({\bf k}'',{\bf k}''')
\nonumber \\
&&+\delta({\bf k}+{\bf k}'')\delta({\bf k}'+{\bf k}''') 
q^{s s''}  ({\bf k},{\bf k}'') q^{s's'''}
({\bf k}',{\bf k}''')
\nonumber \\
&+&\delta({\bf k}+{\bf k}''')\delta({\bf k}'+{\bf k}'')
 q^{s s'''} ({\bf k},{\bf k}''') q^{s' 
s''}({\bf k}',{\bf k}''),
\end{eqnarray} 
where $\left<\dots\right>$ denotes average and the presence of the 
delta function is a direct reflection of spatial homogeneity. Indeed the 
property of spatial homogeneity affords one a way of defining averages, 
which does not depend on the presence of  a joint distribution. We can 
define the average $ \left<\eta({\bf x}) \eta({\bf x}+{\bf r})\right>$ as 
simply an average over the base coordinate, namely 
\begin{equation}
\left<\eta({\bf x})  \eta({\bf x}+{\bf r})\right>=\frac{1}{(2L)^3} 
\int_{-L}^L\eta({\bf x})\eta({\bf x}+{\bf r}) d {\bf x}\ . 
\label{V1}
\end{equation} 
To derive the main results of this paper, it is sufficient to write the  
equations for the second and third order cumulants.  They are 
\begin{eqnarray}
&&\!\!\!\!\!\!\!
\frac{d q^{s s'}_{{\bf k}{\bf k}'}}{d t}=\epsilon P_{00'}
\sum_{s_qs_p}\int 
d {\bf k}_p d {\bf k}_q
L_{{\bf k},{\bf k}_p,{\bf k}_q}^{s,s_p,s_q} q_{{\bf k}'{\bf k}_p {\bf 
k}_q}^{s's_ps_q}
\nonumber \\ &\times&
\exp [i(s_p\omega_p+s_q\omega_q-s\omega_k)t]
\delta({{\bf k}-{\bf  p}-{\bf q}}),\nonumber \\
&&{\bf k}+{\bf k}'=0\ ;
\label{V2}\\
&&\!\!\!\!\!\!\!
\frac{d q^{s s's''}_{{\bf k}{\bf k}'{\bf k}''}}{d t}=\epsilon  P_{00'0''} 
\int d {\bf k}_p d {\bf k}_q 
L_{{\bf k},{\bf k}_p,{\bf k}_q}^{s,s_p,s_q} q_{{\bf k}',{\bf k}'', 
{\bf k}_p, {\bf k}_q}^{s',s'',s_p,s_q} \delta({{\bf k}-{\bf  p}-{\bf q}})
\nonumber \\ &\times&
\exp [i(s_p\omega_p+s_q\omega_g-s\omega_k)t]+ \nonumber \\&&
2  \epsilon  P_{00'0''}\sum_{s_ps_q}L_{{\bf k},-{\bf k}',-{\bf k}''}
^{s,s_p,s_q} q^{s',s_p} _{{\bf k}', -{\bf k}'} 
q^{s'',s_q}_{{\bf k}'',-{\bf k}''}
\nonumber \\&\times&
\exp [i(s_p\omega'+s_q\omega''-s\omega)t]\,,\qquad
{\bf k}+{\bf k}'+{\bf k}''=0 
\label{V4}
\end{eqnarray} 
where the symbol $P_{00'}$ ($P_{0 0' 0''}$) means that we sum over all 
replacements \ \ $0\rightarrow 0', \ \ 0'\rightarrow 0$ ( $0\rightarrow 
0',0'\rightarrow 0'',\ 0''\rightarrow 0,\ 0\rightarrow 0'', 0'\rightarrow 
0,\ \ 0''\rightarrow 0'$). 

The total energy of the system per unite volume can be written as
\begin{eqnarray}
&&\!\!\!\!\!\!
\lim_{r\rightarrow 0} \langle \frac{1}{2}\rho_0 v_j({\bf x}) v_j({\bf 
x}+{\bf r}) 
\nonumber \\
&+&\frac{c^2\rho_0}{\mu}\eta({\bf x}) \eta({\bf x}+{\bf r}) 
+\frac{\rho c^2}{2\mu} (\mu-2)\eta({\bf  x})\eta({\bf x} + {\bf 
r})\rangle
\nonumber \\
&=& \lim_{{\bf r}\rightarrow 0}\sum _{s_1 s_2} \int 
\frac{\rho_0c^2 \epsilon ^2}{2}(1-s_1s_2)q^{s_1s_2}
({\bf k})e^{i{\bf k} {\bf r}}d 
{\bf k}
\nonumber \\
&=&
\lim_{{\bf r}\rightarrow 0}\int\rho_0c^2 \epsilon ^2(q^{+-}({\bf 
k})+q^{-+} ({\bf k}))e^{i{\bf k}{\bf r}} d {\bf k} 
\nonumber \\
&=&  
\int 2\rho_0c^2 \epsilon ^2q^{+-}({\bf k})d {\bf k} 
\end{eqnarray}
since $q^{+-}({\bf k})=q^{-+}(-{\bf k})$. The spectral energy is
therefore $2\rho c^2 \epsilon ^2 q^{+-}({\bf k})$.  For convenience we
denote $q^{+-}({\bf k})$ as $e({\bf k})$.

To leading order in $ \epsilon $, $q^{s s'} ({\bf k}, {\bf k}')$ and
$q^{s s' s''} ({\bf k}, {\bf k}',{\bf k}'')$ (which we may call $q^{s
  s'}_0 ({\bf k}, {\bf k}')$ and $q^{s s' s''}_0 ({\bf k}, {\bf
  k}',{\bf k}'')$) are independent of time. Anticipating, however that
certain parts of the higher order iterates in their asymptotic
expansions may become unbounded, we will allow both $q^{s s'}_0 ({\bf
  k}, {\bf k}')$ and $q^{s s' s''}_0 ({\bf k}, {\bf k}',{\bf k}'')$ to
be slowly varying in time
\begin{eqnarray}
&&\frac{d q_0^{s s'}({\bf k},{\bf k}')}{d t} =  \epsilon ^2 F_2^{s s'}\,,
\nonumber \\ 
&&\frac{d q_0^{s s' s''}({\bf k},{\bf k}',{\bf k}'')}{d t} = 
 \epsilon ^2 F_3^{s 
s' s''}
\end{eqnarray}
and we will choose $F_2$ and $F_3$ to remove those terms with
unbounded growth from the later iteration. We will find that for
$s'=-s$, \ $F_2^{s -s}$ is given by the right-hand side of acoustic KE:
\begin{eqnarray}
F_2^{s s'}= 
   q_0^{s\ s'}({\bf k},{\bf k'})
 \lim_{ \epsilon ^2\rightarrow 0 }\sum_{s_p s_q} \int 
\frac{s_q s_p}{s \omega}\int\left(L_{ {\bf k},{\bf k}_p,{\bf k}_q}
^{s,s_p,s_q}\right)^2
\nonumber\\
\times
q^{s_p,-s_p}({\bf k}_p)\Delta(s_p\omega_p+s_q\omega_q-s\omega)
\delta({\bf k}_p+{\bf k}_q-{\bf k})d {\bf k}_p d {\bf k}_q
\nonumber \\
+\lim_{ \epsilon ^2\rightarrow 0 }\sum_{s_p s_q} \int 
\frac{s_q s_p}{s \omega}\int\left(L_{ {\bf k},{\bf k}_p,{\bf k}_q}
^{s,s_p,s_q}\right)^2
\nonumber\\
\times
 q_0^{s\ s'}({\bf k_q},{-\bf k_q})q^{s_p,-s_p}({\bf k}_p)
\Delta(s_p\omega_p+s_q\omega_q-s\omega) \nonumber \\
\delta({\bf k}_p+{\bf k}_q-{\bf k})d {\bf k}_p d {\bf k}_q
\nonumber 
\end{eqnarray}
and that $F_2^{s s} $ and $F_3^{s s' s''}$ have the form
\begin{eqnarray}
&&i q_0^{s s}({\bf k}, {\bf k}')(\bar \Omega_{\bf k}^s +
 \bar\Omega_{{\bf k}'}^s) 
\label{Newellfinal1} \end{eqnarray} and \begin{eqnarray}
&&i q_0^{s s' s''}({\bf k}, {\bf k}',{\bf k}'')(\bar \Omega_{\bf k}^s 
+ \bar\Omega_{{\bf k}'}^{s'} + \bar\Omega_{{\bf k}''}^{s''}  )
 \label{Newellfinal2}
\end{eqnarray}
respectively.  
It is clear that $\bar\Omega^s_k$ can be interpreted as a complex
frequency modification. Its exact expression is given by
\begin{eqnarray} 
&&\!\!\!\!\!\!\!
\bar\Omega_{\bf k}^s= - 4 i
 \lim_{ \epsilon ^2\rightarrow 0 }\sum_{s_p s_q} \int 
\frac{s_q s_p}{s \omega}\int\left(L_{ {\bf k},{\bf k}_p,{\bf k}_q}
^{s,s_p,s_q}\right)^2
\label{Newellfinal3}\\
&\times&
q^{s_p,-s_p}({\bf k}_p)\Delta(s_p\omega_p+s_q\omega_q-s\omega)
\delta({\bf k}_p+{\bf k}_q-{\bf k})d {\bf k}_p d {\bf k}_q
\nonumber
\end{eqnarray}
and, when calculated out, is precisely equal to 
$s(\omega-c|{\bf k}|)\epsilon ^2$ 
in (\ref{N19}).  Note that in (\ref{Newellfinal3}),
 $t=T\epsilon ^2$ and $T$ is finite.  The 
$\ln(1/ \epsilon ^2)$ coefficient
comes from the term $\ln t$ or 
$\ln (T/ \epsilon ^2)= \ln T + \ln (1/\epsilon ^2)$ 
in the asymptotic expansion. For finite $T$, the dominant part is
$\ln(1/ \epsilon ^2)$.

The perturbations method has the advantage that it is relatively simple
to execute. However, there is no a priori guarantee that terms
appearing later in the formal series cannot have time dependencies
which mean they affect the leading approximations on time scales
comparable to or less than $\epsilon^{-2}$ (e.g. a term 
$\epsilon^4t^3$
should be accounted for before the term $\epsilon^2t$). To check this,
one must have a systematic approach for exploring all orders in the
formal perturbation series and removing (renormalizing) in groups those
resonances which make their cumulative effects at time scales
   $\epsilon^{-N}(\ln(\frac{1}{\epsilon})^{-M})$, 
$N,M=1,2,3, ...$. 
The diagram approach, which requires some familiarity to
execute, is designed to do this and, both for completeness  and the
fact that we will have to proceed beyond the one-loop approximation to
resolve the questions of the angular redistribution of spectral
energy, we include it here.

\section{Diagrammatic Approach to Acoustic Turbulence} 
\subsection{Objects of Diagrammatic Technique}
Let us define the ``bare'' Green's function  of Eq. (\ref{A17}) as 
\begin{equation} 
G_0({\bf k})=\frac{1}{\omega-c k+i 0}\ .
\end{equation} 
One may see from (\ref{A17}) that this function describes the response
of the system of noninteracting acoustic waves on some external force.
We added in the denominator the term $+i 0$ by requirement of
causality. We remark that causality (the arrow of time) is introduces
in the perturbation approach by the limit $t\to\infty$ and the fact
that ${\rm sgn}t$ appears in (\ref{N8}).  Next we introduce the
``dressed'' Green-function which is the response of interacting wave
systems on this force:
\begin{equation} 
(2\pi)^4 G({\bf k},\omega)\delta({\bf k}-{\bf k}')\delta(\omega-\omega')
=\left<\frac{\delta b({\bf k},\omega)}
{\delta f({\bf k}',\omega')}\right>\ .
\label{E2} 
\end{equation}
We will be interested also in the double correlation function $n({\bf
  k},\omega)$ of the acoustic field $b,b^*$
\begin{equation} 
(2\pi)^4 n({\bf k},\omega)\delta({\bf k}-{\bf k}')\delta(\omega-\omega')
    =\left<b({\bf k},\omega)b^*({\bf k}',\omega')\right>  \ .
\label{E3}
\end{equation}
The simultaneous double correlator of the acoustic field $n({\bf k})$
is determined by
\begin{equation} 
(2\pi)^3 n({\bf k}) \delta({\bf k}-{\bf k}')= 
     \left<b({\bf k},t)b^*({\bf k}`,t)\right>\  . 
\label{Eo4} 
\end{equation}
This is related to the different-time correlators in the $\omega$
representation $n({\bf k},\omega)$ as follows:
\begin{equation} 
n({\bf k})=\int n({\bf k},\omega) \frac{d \omega}{2\pi}\ .  
\label{Eo5} 
\end{equation}

The Green's and correlation functions together with the bare vertex
$V({\bf k},{\bf q},{\bf p})$ (\ref{A13}) are the basic objects of
diagrammatic perturbation approach which we are going to use (see Fig
1a).
\narrowtext
\vskip .5cm
\begin{figure}
 \epsfxsize=8.6truecm 
 \vspace{.5cm}
 \caption{Panel (a): Basic objects of diagrammatic pertubation approach.
Panel (b): First terms in the expansion of mass operator  $\Sigma{({\bf 
k},\omega )}$.} 
\label{Fig1} \end{figure} 
\begin{figure}
 \epsfxsize=8.6truecm 
 \vspace{.5cm}
 \caption{Diagrams (a) from Fig.1 with specified directions of arrows.}
 \label{Fig2}  
 \end{figure} 
\subsection{The Dyson-Wyld equations}
In the diagrammatic series for the Green's function one may perform
the partial Dyson's summation over one-particle irreducible
diagrams. This results in the Dyson equation for the Green's functions:
\begin{equation}
G({\bf k},\omega) = \frac{1}{\omega-\omega_0(k)
+i0 -\Sigma({\bf k},\omega)}\label{C1}
\end{equation}
where the ``mass operator'' $\Sigma({\bf k},\omega)$ gives the
nonlinear correction to the complex frequency $\omega_0({\bf k})+i 0$
due to the interaction (\ref{A9}). This is an infinite series with
respect to the bare amplitude $V({\bf k},{\bf q},{\bf p})$
(\ref{A13}), dressed Green's function (\ref{E2}) and double
correlation function $n({\bf k},\omega)\ $ (\ref{E3}).  All of the
contributions of the second and fourth order in $V$ are shown on Fig
1(b).

We have not specified the direction of arrows on Fig 1(b); each
diagram should be interpreted as a sum  of diagrams with all possible
directions of arrows compatible with vortex $V({\bf k},{\bf q},{\bf
p})$, describing the  three-wave processes $1\leftrightarrow 2$.  For
example, diagram (a) on Fig 1(b) corresponds to three diagrams shown
on Fig 2. The diagram (a4) on Fig 2 describes the nonresonant process
$0\leftrightarrow 3$ which is not essential for our consideration.

With the help of the similar Dyson's summing of one-particle
irreducible diagrams, one can derive Wyld's equation for $n({\bf
k},\omega)$:
\begin{equation}
n({\bf k},\omega) = |G({\bf k},\omega)|^2\left[D({\bf k},\omega
)+\Phi({\bf k},\omega)\right] \ .
 \label{C2}
\end{equation}
Here $D({\bf k},\omega)$ is the correlation function of white noise,
\FL
\begin{equation}
(2\pi)^4 D({\bf k},\omega)\delta({\bf k}-{\bf k}')\delta(\omega-\omega')
 = \left<f({k \omega}) f^*({k'\omega'})\right>,
\label{C3}
\end{equation}
and the mass operator $\Phi({\bf k},\omega)$ describes the nonlinear
corrections to $D({\bf k},\omega)$.  This is an infinite series with
respect to the same objects $G({\bf k},\omega)$,$n({\bf k},\omega)$
and $V({\bf k},{\bf q},{\bf p})$.  All diagrams of the second and
fourth order are shown on Fig 3(a).

We also have not specified arrow directions in the diagrams for
$\Sigma({\bf k},\omega)$ and $\Phi({\bf k},\omega)$.  In complete
analogy with diagrams for $G({\bf k},\omega)$ one diagram on Fig 3a
corresponds to {\it two} diagrams (a1) and (a2) on Fig 3b.  All the
rest diagrams for $\Phi({\bf k},\omega)$ reproduces in the same way -
one chooses all possible directions of arrows and discards those which
incompatible with definition of vertex $V$ (see Fig 1a).
\vskip 1cm
\begin{figure}
 \epsfxsize=8.6truecm 
 \vspace{.5cm}
 \caption{First terms in the diadrammatic pertubation expansion for 
mass operator $\Psi{({\bf k},\omega )}$.}
 \label{Fig3}  
 \end{figure} 
\vskip 1cm

\subsection{One-pole approximation}
\subsubsection{The Green's function}
We have assumed from the  beginning, that the wave  amplitude is small. 
Therefore, 
\begin{equation} 
\Sigma({\bf k},\omega)\ll\omega_0(k)\ .
\label{D1}
\end{equation}
As a result the Green's function has a sharp peak in the vicinity of 
$\omega=c k$ and one may (as a first step in the  analysis) neglect 
the $\omega$-dependence of $\Sigma({\bf k},\omega)$ and put 
\begin{equation}
 \Sigma({\bf k},\omega)\simeq\Sigma({\bf k},\omega\simeq c {\bf k}) \ .
\label{D11}
\end{equation}
The validity of this assumption will be checked later.  Under this 
assumption the Green's function (\ref{E2}) has a simple one-pole 
structure:
\begin{equation}
 \tilde G({\bf k},\omega) = \frac{1}{\omega-\omega({\bf k})
+i \gamma({\bf k})}\,,
\label{D2}
\end{equation}
where  
\begin{eqnarray} \omega({\bf k})&=&\omega_0({\bf k})
+{\rm Re}\Sigma({\bf k},\omega_*)\,,
\label{D3}\\
 \gamma({\bf k})&=& -{\rm Im}\Sigma({\bf k},\omega_*)\ .
\label{D4}
\end{eqnarray}
Now we have to decide how to choose $\omega_*$ ``in the best way''.
The simplest way is to put $\omega_*=\omega_0({\bf k})=c k$, as it was
stated in (\ref{D11}). As a next step we can take ``more accurate''
expression $\omega_*=\omega({\bf k})$, i.e. to take into account the
real part of correction to $\omega_0({\bf k})$. But later we will see,
that better choice is
\begin{equation}
 \omega_*=\omega({\bf k})+i \gamma({\bf k})
 \label{D15}
\end{equation}
which is consistent with the position of the pole of $\tilde G_*({\bf
  k},\omega)$. We will show that this choice is self consistent while
deriving the balance
equation in section 5.3. 
\subsubsection{The double correlation function}
The same type of approximation may be performed for the correlation
function. Namely in the Wyld equation (\ref{C2}) one may replace
$G({\bf k},\omega)$ by $\tilde G({\bf k},\omega)$ and to neglect the
$\omega$ dependence of $\Phi({\bf k},\omega)$ by putting $\Phi({\bf
  k},\omega)\rightarrow\tilde\Phi(k)=\Phi({\bf k},\omega_*)$, or
\begin{equation}                                            
\tilde n({\bf k},\omega)=
|\tilde G({\bf k},\omega)|^2\left[D(k)+\tilde\Phi(k)\right]\,,
\label{D5}
\end{equation}
We will call this {\sl one-pole approximation for the correlation 
function}.
\subsection{One-loop approximation}
Let us begin our treatment with the simple one-loop (or direct
interaction) approximation for mass operators $\Sigma$ and $\Phi$.
This approximation corresponds to taking into account just the second
order (in bare vertex $V$ (\ref{A13})) diagrams for the mass operators
$\Sigma$ and $\Phi$.  Two loop approximation will be considered in
Appendix C. We will estimate two-loops diagrams and we will show, that
some of them gives the same order contribution to $\gamma_k$ as
one-loop diagrams.  Therefore, one loop approximation is an
uncontrolled approximation, but we believe, that it gives
qualitatively correct results.  Note that these diagrams include the
dressed Green's function in contrast to the approximation of kinetic
equation which is nothing but one-loop approximation with the bare
Green's function inside.  We will see later that this difference is
very important in particular case of acoustic turbulence. The KE for
waves with linear dispersion law forbids the angular evolution of
energy because conservation laws of energy and momentum allow
interaction only for waves with parallel wave vectors.  In the
one-loop approximation with dressed Green's function, the conservation
laws $\omega({\bf k})\pm\omega({\bf k}_1) =\omega({\bf k}\pm{\bf
k}_1)$ are satisfied with some accuracy [of the order of $ \gamma({\bf
k})$]. As a result, there exists a cone of allowed angles between
${\bf k}$ and ${\bf k}_1$ in which interactions are allowed.
Therefore one has to expect some angle evolution of wave packages
within this approximation.  Combining (\ref{D2}) with (\ref{D5}) one
has the following expression:
\begin{equation}
n({\bf k},\omega)=\frac{2 \gamma({\bf k})\tilde n({\bf k})}
{[\omega-\omega({\bf k})]^2+ \gamma^2(k)} \ .
\label{D6}
\end{equation}
\subsubsection{Calculations of $\Sigma({\bf k},\omega)$}
In the one-loop approximation expression for $\Sigma({\bf k},\omega)$
has the form
\begin{equation} 
\Sigma({\bf k},\omega)= \Sigma_{a1}({\bf k},\omega)
+ \Sigma_{a2}({\bf k},\omega) + \Sigma_{a3}({\bf k},\omega) \,,
\end{equation}
where $\Sigma_j({\bf k},\omega)$ is given by (\ref{C4}-\ref{C6}).  Our
goal here is to analyze these expressions in one-pole approximation,
by substituting in it ``one-pole'' $n({\bf k},\omega)$ and $G({\bf
k},\omega)$ from (\ref{D2}) and (\ref{D6}). In the resulting
expression one can perform the integration over $\omega$ analytically.
The result is
\begin{eqnarray}
&&\Sigma({\bf k},\omega)=\int  \frac{d^3 k_1 d^3 k_2}{(2\pi)^3}
\label{Eo1}  \\
&\times&
\Bigg( \frac{|V({\bf k}_2,{\bf k},{\bf k}_1)|^2 
\delta({\bf k}+{\bf k}_1-{\bf k}_2)
\big[n({\bf k}_1)-n({\bf k}_2)\big]}
     {\omega+\omega({\bf k}_1)-\omega({\bf k}_2)
+i ( \gamma_1+ \gamma_2)}
\nonumber \\ &+&
\frac{|V({\bf k}_0,{\bf k}_1,{\bf k}_2)|^2
 \delta({\bf k}-{\bf k}_1-{\bf k}_2)n({\bf k}_2)}
     {\omega-\omega({\bf k}_1)-\omega({\bf k}_2)+i
 ( \gamma_1+ \gamma_2)}\Bigg) \ .
\nonumber
\end{eqnarray}
Next we introduce $\Sigma({\bf k})=\Sigma({\bf k},\omega_*)$ , with
$\omega_*$ given by (\ref{D15}) and consider (\ref{Eo1}) in the limit
of small $\gamma$, which allows us to perform analytically
integrations over perpendicular components of wavevectors. The result
for the damping frequency $\gamma(k)$ may be represented in the
following form (for details see Appendix B):
\begin{eqnarray} 
 \gamma(k)=\frac{A^2 k^2}{4 \pi c }\int_{1/L}^\infty
{n(q) q^2 d q}
\simeq 
\frac{A^2 k^2 }{4\pi c } N(\Omega) \ .
\label {E11}
\end{eqnarray}
We introduced here cut-off for small $k$ at $1/L$, where $L$ is the
size of the box.  We also introduced ``the density of the number of
particles'' $N(\Omega)$ in the solid angle according to
\begin{equation} 
N(\Omega)=\int k^2  n({\bf k}) d k \ ,
\label{E11o1}
\end{equation}
such that the total number of particles 
\begin{equation} 
N=\int N(\Omega)d\Omega\ . 
\label{E11o2}
\end{equation}
After substituting $A$ from (\ref{E82}), one has the following
estimate for $ \gamma({\bf k})$:
\begin{equation} 
 \gamma({\bf k})\simeq k^2 N(\Omega)/\rho_0\, , 
\label{E11o3}
\end{equation}

Consider now $\Sigma'({\bf k})\equiv {\rm Re \Sigma({\bf k})}$. It follows
from (\ref{E9}) that 
\begin{eqnarray} 
\Sigma({\bf k})&=&\frac{A^2}{4 \pi ^2  c }
\int d q \int _0 d y q^2 n(q) \frac{y}{y^2+\Gamma_{k12}^2}
\label{E13}\\
&&\times
\left[ (k^2 + 2 k q + q^2) - (k^2 - 2 k q + q^2 ) \right]
\nonumber \\
&\simeq &\frac{A^2 k } { \pi^2  c^2}\int d q \int_0^{y_{\rm max}}
\frac{y d y }{y^2 + \Gamma_{k12}^2}
\left[ c q^3 n(q)  \right] \,, 
\nonumber
\end{eqnarray}
where $\Gamma_{k12}=\gamma (k)+\gamma (k_1)+\gamma (k_2)$ is the
``triad interaction'' frequency.  One may evaluate the integral with
respect to $y$ as
\begin{equation} 
\L(q)=\ln{\frac{y_{\rm max}}{\Gamma_{kkq}}}
\simeq \ln{\frac{c k^2}{q \gamma({\bf k})}}. \end{equation}
After substituting $ \gamma({\bf k})$ from (\ref{E11o3}),  one has
\begin{equation} 
\L(q)\propto \ln{\rho_0\c/q N(\Omega)} 
\label{65}
\end{equation}
The main contribution to the integral (\ref{E13}) over $q$ comes from
the infrared region $q\simeq 1/L$. It gives the estimate,
\begin{equation}
\Sigma'({\bf k}) = \frac{A^2 k}{\pi^2 c^2}\L E(\Omega)
\label{E15}
\end{equation}
where we have defined the density of the wave energy in solid angle as
\begin{equation} 
E(\Omega)= \int \omega_0(k)n(k) k^2 d k\ .
\label{E16}
\end{equation}
This value relates to $N(\Omega)$ as follows:
\begin{equation} 
E(\Omega)\simeq \frac{ c }{L}N(\Omega) \ . 
\end{equation}
Equation (\ref{E15}) together with the expression (\ref{E82}) for $A$ may
be written as
\begin{equation} 
\Sigma'({\bf k})\simeq c k \epsilon \ln{1/\epsilon}\ , 
\label{E20}
\end{equation}
where 
\begin{equation}
\epsilon\simeq E(\Omega)/\rho_0 c^2 \ ,
\label{E21}
\end{equation}
is the dimensionless parameter of nonlinearity, the ratio of energy of
acoustic turbulence and the density of thermal energy of media $\rho_0
c^2 \simeq \tilde n T$, where $\tilde n$ is the concentration of
atoms.

Equation (\ref{E11o3}) for $ \gamma({\bf k})$ may be written 
in a similar form 
\begin{equation}
 \gamma({\bf k})\simeq c k (k L) \epsilon\ . 
\label {E23}
\end{equation}
One can see that 
\begin{equation}
\frac{ \gamma({\bf k})}{\Sigma'({\bf k})}\propto \frac{k L}
{\ln{1/\epsilon}}\ .
\label {E24}
\end{equation}
It means that for large enough inertial interval
\begin{equation}
 \gamma({\bf k}) \gg\Sigma'({\bf k}) 
\label {E25}
\end{equation}
and one may neglect the nonlinear corrections $\Sigma'({\bf k})$ to
the frequency with respect to the damping of the waves $ \gamma({\bf
k})$.  That shows that our above calculations of $\Sigma({\bf k})$ is
self-consistent. Later we also will take into account only damping $
\gamma({\bf k})$ in the expressions for the Green's functions taking
$\omega(k)=\omega_0(k)=c k$.
\subsubsection{Calculations of $\Phi({\bf k},\omega)$.}
In the one-loop approximation expression for $\Phi({\bf k},\omega)$
has the form (\ref{C7}). After substitution of $n({\bf k},\omega)$ in
the one pole approximation (\ref{D6}) one may perform analytically
integration over frequencies:
\begin{eqnarray}
&&\Phi({\bf k},\omega)=\int \frac{d^3 k_1 d^3 k_2}{(2\pi)^3}
 n({\bf k}_1)n({\bf k}_2)
\label{F1}\\
&\times&
\Bigg[ \frac 
{|V({\bf k},{\bf k}_1,{\bf k}_2)|^2 ( \gamma({\bf k}_1)+ \gamma({\bf
k}_2))\delta({\bf k}-{\bf k}_1-{\bf k}_2)} {\left[\omega-\omega({\bf
k}_1)-\omega({\bf k}_2)\right]^2 + \left[ \gamma({\bf k}_1)+
\gamma({\bf k}_2)\right]^2}\nonumber \\ &+& \frac {|V({\bf k}_2,{\bf
k}_1,{\bf k})|^2 ( \gamma({\bf k}_1)+ \gamma({\bf k}_2))\delta({\bf
k}+{\bf k}_1-{\bf k}_2)} {\left[\omega+\omega({\bf k}_1)-\omega({\bf
k}_2)\right]^2 +
\left[ \gamma({\bf k}_1)+ \gamma({\bf k}_2)\right]^2} 
      \Bigg] \ .
\nonumber
\end{eqnarray}
We will  analyze this expression  in the next section.
\subsubsection{Balance Equation}
Consider the  Dyson-Wyld equations (\ref{C1}) and (\ref{C2}) in the
inertial interval, where one can neglect $ \gamma_0({\bf k}) $ in
comparison with ${\rm Im} \Sigma({\bf k},\omega)$ and $D({\bf k}) $ in
comparison with $\Phi({\bf k},\omega)$:
\begin{eqnarray}  G({\bf k},\omega) &=&\frac{1}{\omega-\omega_0({\bf k})
-\Sigma({\bf k},\omega)} \,, 
\label {alpha1}\\
n({\bf k},\omega)&=&|G({\bf k},\omega)|^2 \Phi({\bf k},\omega) \ .
 \label {alpha2}
\end{eqnarray}
It follows from (\ref{C1}) that 
\begin{eqnarray}  {\rm Im}G({\bf k},\omega)=|G({\bf k},\omega)|^2{\rm Im}
\Sigma({\bf k},\omega).\label{alpha3} \end{eqnarray}
By comparing (\ref{alpha2}) with
 (\ref{alpha3}) one may 
see that the following combination 
\begin{equation}
L({\bf k},\omega)\equiv    \Phi({\bf k},\omega) {\rm Im}
 G({\bf k},\omega) -n({\bf k},\omega){\rm Im}\Sigma({\bf k},\omega)
\label{alpha4}
\end{equation}
is equal to zero.  In particular
\begin{equation} 
L({\bf k})\equiv\int L({\bf k},\omega) \frac{d\omega}{2\pi}=0\ .
\label{alpha}  
\end{equation}
Together with (\ref{alpha4}) it gives
\begin{equation}
{\rm Im} \int \frac{d\omega}{2\pi}
\left[ \Phi({\bf k},\omega) G({\bf k},\omega)
 - n({\bf k},\omega)\Sigma({\bf k},\omega) \right]=0\ .
\label{alpha6} 
\end{equation}
Let us compute now the first term in (\ref{alpha6}). By substitution
Eq.  (\ref{F1}) for $\Phi({\bf k},\omega)$ and Eq. (\ref{D2}) and
integration over $\omega$ one has
\begin{eqnarray}
&&\int \frac{d\omega}{2\pi}G({\bf k},\omega)
\Phi({\bf k},\omega)=\int \frac{d{\bf k}_1 d {\bf k}_2}{(2\pi)^3} 
n({\bf k}_1)n({\bf k}_2)
\nonumber \\&\times&
\Bigg[ \frac{1}{2} 
        \frac{|V({\bf k},{\bf k}_1,{\bf k}_2)|^2
 \delta({\bf k}-{\bf k}_1-{\bf k}_2)}
             {\omega_0({\bf k})-\omega_0({\bf k}_1
)-\omega_0({\bf k}_2)-i\Gamma_{k12}}\nonumber \\
&+&\frac{|V({\bf k}_2,{\bf k}_1,{\bf k})|^2
 \delta({\bf k}+{\bf k}_1-{\bf k}_2)}
{\omega_0({\bf k})+\omega_0({\bf k}_1)-\omega_0({\bf k}_2)
-i\Gamma_{k12}} \Bigg]
\ .
\label{F2}
\end{eqnarray} 

Next we will perform integration over $\omega$ in (\ref{alpha6}).
Remember $\Sigma({\bf k},\omega)$
is analytical function in the upper half plane of $\omega$
while $n({\bf k},\omega)$ has one pole there. Therefore
\begin{equation} {\rm Im} \int \frac{d\omega}{2\pi} n({\bf k},
\omega)\Sigma({\bf k},\omega)=
     n({\bf k}){\rm Im}\Sigma({\bf k},\omega_*)
\label{alpha7}
\end{equation}
where $\omega_*$ is given by (\ref{D5}).
This is the justification of our choice 
$\omega_*$. 

Now let us put everything together to obtain 
\begin{eqnarray}
&&\!\!\!\!\!\!\!
0=L({\bf k})=\int\frac{d{\bf k}_1 d{\bf k}_2}{(2\pi)^3}\Gamma_{k12}
\Bigg\{
\delta({\bf k}-{\bf k}_1-{\bf k}_2)
\label{F3}\\
&\times&\frac{1}{2}\frac{|V({\bf k},{\bf k}_1,{\bf k}_2)|^2 
  \left[n({\bf k}_1)n({\bf k}_2)-n({\bf k})[n({\bf k}_1
)+n({\bf k}_2)]\right]}
   {(\omega_0({\bf k})-\omega_0({\bf k}_1)-\omega_0({\bf k}_2))^2
+\Gamma^2_{k12}}
\nonumber \\
&+&\delta({\bf k}+{\bf k}_1-{\bf k}_2)   
\nonumber \\
&\times&
\frac{|V({\bf k}_2,{\bf k}_1,{\bf k})|^2 
  \left[n({\bf k}_2)[n({\bf k}_1)+n({\bf k})]
 -n({\bf k})n({\bf k}_1)\right]}
   {(\omega_0({\bf k})+\omega_0({\bf k}_1
)-\omega_0({\bf k}_2))^2+\Gamma^2_{k12}}
\Bigg\} \ .
\nonumber
\end{eqnarray}
This is the main result of the diagrammatic approach: the {\it balance
  equation for stationary in time acoustic turbulence}.  In
nonstationary case one can get similarly the {\it generalized kinetic
  equation} in the form
\begin{equation}
{\partial n({\bf k}, t)\over \partial t}=L({\bf k}, t)
\label{GKE}
\end{equation}
where $L({\bf k}, t)$ is given by Eq. (\ref{F3}) with correlator
depending on time $n({\bf k}_j)\to n({\bf k}_j,t)$.  In the limit $
\gamma ({\bf k}) \to 0$ this expression turns into well known (cf.
\cite{ZLF}) collision integral for 3-wave kinetic equation
\begin{eqnarray}
&&\!\!\!\!\!\!  {\cal S}t\{n({\bf k}, t) \}=\lim_{\Gamma_{k12}\to 0}
L({\bf k}) =2\pi \int\frac{d{\bf k}_1 d{\bf k}_2}{(2\pi)^3}
\nonumber\\
&\times& \frac{1}{2}\delta({\bf k}-{\bf k}_1-{\bf k}_2) |V({\bf
k},{\bf k}_1,{\bf k}_2)|^2 \nonumber\\ && \left\{n({\bf k}_1)n({\bf
k}_2)-n({\bf k})[n({\bf k}_1)+n({\bf k}_2)]\right\}
\nonumber\\
 &\times& \delta [\omega_0({\bf k})-\omega_0({\bf k}_1)-\omega_0({\bf
k}_2)]
\nonumber\\
&+& \delta({\bf k}+{\bf k}_1-{\bf k}_2) |V({\bf k}_2,{\bf k}_1,{\bf
k})|^2 \nonumber\\&& \left\{n({\bf k}_2)[n({\bf k}_1)+n({\bf k})]
-n({\bf k})n({\bf k}_1)\right\}
\nonumber\\
&\times & 
\delta [\omega_0({\bf k})+\omega_0({\bf k}_1)-\omega_0({\bf k}_2)]   \ .
\label{coint}
\end{eqnarray}
We see that the generalized kinetic equation differs from the well
known collision term in the three wave kinetic equation by replacing
$\delta$-functions on the corresponding Lorenz function with the width
of $\Gamma_{k12}$-triad interaction frequency.

\section{Conclusion}
In the present paper we have begun to develop a consistent statistical
description of acoustic turbulence based both on the long-time
asymptotic analyses (Section 3) and on the perturbation diagrammatic
approach (Section 4). The first approach is more straightforward. The
diagrammatic approach provides a systematic way of analyzing higher
order terms in the perturbation theory.

Our main result is that the nonlinear corrections to the frequency is
much smaller than the nonlinear damping of the waves. We find also the
balance equation (\ref{F3}) which generalizes the simple kinetic
equation for acoustic waves.  One can show that the balance equation
(\ref{F3}) has the same isotropic solution (Zakharov-Sagdeev spectrum)
as the kinetic equation.  However the kinetic equation for acoustic
turbulence does not describe the angle evolution of turbulence: any
arbitrary angle distribution is the solution of KE. In contrast, our
balance equation (\ref{F3}) contains terms which describe an angular
redistribution of the energy because of the non-zero value of the
interaction cone, which is proportional to $\Gamma_{k12}$.  But we
have yet to show that this expression contains all such terms to this
order.

One may imagine three very different ways of the angle evolution of
anisotropic acoustic turbulence. The first one is a tendency to form
very narrow beams with the characteristic width of about interaction
angle. The second one is an approach to isotropy downstream to the
large wave vectors. The last possibility is to form a beam with a
characteristic width of about unity, exactly like it happens in the
turbulence of waves with weak dispersion \cite{LF}. Another important
question is: does the spectra of acoustic turbulence depend on the
features of pumping or they are universal (independent of details of
energy influx)? We intend to answer these questions (in the framework
approximations we made in that paper) in our next project.

It is an exciting challenge to try to go beyond the approximations
made here in order to understand whether the scaling index of the
interaction vertex in the system of acoustic waves in two- and
three-dimensional media must be renormalized or not.
\acknowledgements We acknowledge full spectrum of grants.  V.L.
acknowledges the hospitality of the Arizona Center for Mathematical
Studies at the University of Arizona, where a portion of this
manuscript was written.

{ \appendix
\section{
Rules for writing and reading of diagrams for mass operators } Here we
state without proof the set of rules for writing down diagrammatic
series:
\begin{enumerate}
\item 
In order to write down all diagrams for $\Sigma $ and $\Phi$ of 2n
order in vertices, one should draw 2n vertices and connect them with
each other by lines $n$ and $G$ in all possible ways.  Two ends must
be left free.  If both ends are straight, we shall get a diagram for
$\Phi({\bf k},\omega)$; if one of them is wavy, this will be a diagram
for $\Sigma({\bf k},\omega)$.
\item
The diagrams for $\Phi $ and $\Sigma $ containing closed loops in GF
are absent.  This follows from the fact that the Wyld's DT appears
from glued trees.
\item
There is no mass operator with two wavy ends in DT.
\item
In the diagrams for $\Phi$ (for $\Sigma $) one can pass from every
vertex along the $G$ lines to the entrance and to the exit in a single
way.
\item In every diagram for $\Sigma $ there is a single root linking
the entrance and exit along the $G$ lines -- the backbone of the
diagram.  The rest $G$ lines of the diagrams may be called the rips.
\item
The diagrams for $\Phi$ contain the basic cross section in which they
may be cut in a single way into two parts only at lines $n({\bf
k},\omega)$.
\item
Every $V$ vertex is entered by one arrow and excited by two.  The
$V^*$ vertex is entered by two arrows and exited by one.
\end{enumerate}

One can show (see \cite{W961}) that rules (3-7) follows from (1-2).

The rules of reading diagrams are the follows:

\begin {enumerate}
\item
Write down product of DT objects (double correlator, Green function or
vertex) (with corresponding arguments) corresponding to each element
of the diagram.
\item 
 Write down delta-functions in 4-momenta for $2n-1$ vertices in such a
way, that the sum of entering 4-momenta is equal to the sum of
exciting ones. One of vertices (for example the one corresponding to
the end of the diagram) does not contain the delta function.
\item 
Perform integration along all internal lines of diagram : $d i = d k_i
/(2\pi)^d d\omega_i/(2\pi)$ where $d$ is space dimension.
\item 
Then you have to multiply diagram by $(2\pi)^{(d+1)}$.
\item 
To multiply diagram by $1/p$ where $p$ is the number of elements in
its symmetry group.
\end{enumerate}
For example diagrams (a1), (a2) and (a3) correspond to the following
analytical expressions: q\begin{eqnarray} &&\Sigma_{a1}({\bf
k},\omega) = \int \frac{d^3 k_1 d^3 k_2}{(2\pi)^3} \frac{d\omega_1
d\omega_2}{2\pi} \delta({\bf k}+ {\bf k}_1 - {\bf k}_2 ) \nonumber
\\&\times& \delta(\omega+\omega_1-\omega_2)|V({\bf k}_2,{\bf k},{\bf
k}_1)|^2 G_2 n_1,\label{C4}\\
&& \Sigma_{a2}({\bf k},\omega) = \int \frac{d^3 k_1 d^3
k_2}{(2\pi)^3}\frac{d\omega_1 d\omega_2}{2\pi} \delta({\bf k}+ {\bf
k}_1 - {\bf k}_2 ) \nonumber \\&\times&
\delta(\omega+\omega_1-\omega_2)|V({\bf k}_2,{\bf k},{\bf k}_1)|^2
G_1^* n_2,\label{C5}\\
&& \Sigma_{a3}({\bf k},\omega) = \int \frac{d^3 k_1 d^3
k_2}{(2\pi)^3}\frac{d\omega_1 d\omega_2}{2\pi} \delta({\bf k}- {\bf
k}_1 - {\bf k} _2 ) \nonumber \\&\times&
\delta(\omega-\omega_1-\omega_2) |V({\bf k},{\bf k}_1,{\bf k}_2)|^2
G_1 n_2,\label{C6}
\end{eqnarray}
We defined here the following shorthand notation, $G_j=G({\bf
k}_j,\omega_j); n_i=n_i({\bf k}_i,\omega_i)$ In the same way one can
find analytical expressions for $\Phi_a({\bf k},\omega)$:
\begin{eqnarray}
&&\Phi({\bf k},\omega)=\int \frac{d^3 k_1 d^3
k_2}{(2\pi)^3}\frac{d\omega_1 d\omega_2}{2\pi}[ \frac{1}{2}|V({\bf
k},{\bf k}_1,{\bf k}_2)|^2 \nonumber \\ & \times & n_1 n_2 \delta({\bf
k}-{\bf k}_1-{\bf k}_2) \delta(\omega-\omega_1-\omega_2)
\label{C7} \\ 
&+&|V({\bf k}_2,{\bf k}_1,{\bf k})|^2 n_1 n_2 \delta({\bf k}+{\bf
k}_1-{\bf k}_2) \delta(\omega+\omega_1-\omega_2) ] \ .  \nonumber
\end{eqnarray}
Analytical expressions for s the 4-th order diagrams (two-loop
diagrams) will be shown in Appendix C.
\section{Calculation of $\Sigma({\bf k},\omega)$ -details}
Let us start from (\ref{Eo1}) and 
introduce $\Sigma({\bf k})=\Sigma({\bf k},\omega_*)$ with
$\omega_*$ given by (\ref{D15}):
\begin{eqnarray}
&&\Sigma({\bf k})= \int  \frac{d^3 k_1 d^3 k_2}{(2\pi)^3}
\label{EO}\\
&\times& \Bigg(
\frac{|V({\bf k}_2,{\bf k},{\bf k}_1)|^2
 \delta({\bf k}+{\bf k}_1-{\bf k}_2)
\big[n({\bf k}_1)-n({\bf k}_2)\big]}
     {\omega({\bf k})+\omega({\bf k}_1)
-\omega({\bf k}_2)+i\Gamma_{k12} }
\nonumber \\ &+&
\frac{|V({\bf k}_0,{\bf k}_1,{\bf k}_2)|^2 
\delta({\bf k}-{\bf k}_1-{\bf k}_2)n({\bf k}_2)}
     {\omega({\bf k})-\omega({\bf k}_1)-\omega({\bf
k}_2)+i\Gamma_{k12} }\Bigg)
\nonumber
\end{eqnarray}
where \begin{equation}
 \Gamma_{k12}= \gamma({\bf k})+ \gamma({\bf
  k}_1)+ \gamma({\bf k}_2)\label{E1}
\end{equation}
is ``triad-interaction'' frequency and $1/\Gamma_{k12}$ is triad
interaction time.  One can consider (\ref{EO} - \ref{E1}) as an
integral equation for the damping of wave $ \gamma({\bf k})=-{\rm
  Im}\Sigma({\bf k})$ and for the frequency $\omega({\bf
  k})=\omega_0({\bf k})+{\rm Re}\Sigma({\bf k})$.

First we consider these equations in the limit of weak interaction
where, $\Gamma\to 0$, and the main contribution to the first term in
(\ref{EO}) comes from the region where
\begin{equation} 
\omega({\bf k})+\omega({\bf k}_1)=\omega({\bf k}_2), \qquad
    {\bf k}+{\bf k}_1={\bf k}_2 \ .
\label{E2o1}
 \end{equation}
 These are conservation laws for 3-wave confluence processes $0+1\to
 2$.  The main contribution for the second term in (\ref{EO}) comes
 from the region
\begin{equation}
\omega({\bf k})=\omega({\bf k}_1)+\omega({\bf k}_2),
 \qquad {\bf k}={\bf k}_1+{\bf k}_2 \ .  
\label{E3o1} 
\end{equation}
These are conservation laws for decays processes $0\to 1+2$.  For weak
interaction one may replace in (\ref{E2}) and (\ref{E3}) $\ \ 
\omega({\bf k})$ on $\omega_0({\bf k})=c k$. Than it follows from
(\ref{E2}) and (\ref{E3}) that ${\bf k}_1 \parallel {\bf k}_2
\parallel {\bf k}$ with ${\bf k}_1,{\bf k}_2$ directed along ${\bf
  k}$. This fact makes it natural to introduce in integrals (\ref{EO})
new variables: scale positive variable $q>0$ and two-dimensional
vector $\bf \kappa$ such that
\begin{equation}
 {\bf k}_1=q {\bf k} /k +{\bf \kappa }\,,
 \qquad {\bbox \kappa }\perp{\bf k}\ .  
\label{E4}\
\end{equation}
In the first term of (\ref{EO})
\begin{equation} {\bf k}_2=(k+q){\bf k} /k +{\bf \kappa }, 
\qquad 0\le q \ .
\label{E5}\end{equation}
In the second term
\begin{equation} {\bf k}_2=(k-q){\bf k} /k -{\bf \kappa }, 
\qquad  0\le q\le k \ . 
\label{E6}\end{equation}
For $\kappa\ll k$ the denominators in integrals (\ref{EO}) strongly
depend on $\kappa$. Indeed:
\begin{eqnarray}
 \omega_0({\bf k})+\omega({\bf k}_1)-\omega(|{\bf k}+{\bf k}_1|)
&=&c k \frac{\kappa^2}{2q(k+q)}\,,
\label{E7}\\
\omega_0({\bf k})-\omega({\bf k}_1)-\omega(|{\bf k}-{\bf k}_1|)
 &=&-c k \frac{\kappa^2}   {2q(k-q)} \ .
\label{E8}
\end{eqnarray}
This allows to neglect $\kappa$ dependence of interaction $V({\bf
k},{\bf q},{\bf p})$ and correlation $n({\bf k}_i)$ in numerator of
(\ref{EO}) for estimation. The result is
\begin{eqnarray}
\Sigma({\bf k})&=&\frac{A^2k}{8\pi^2}\int_0^{k^2}d\kappa^2
\Bigg[ \int_0^\infty d q \frac{q(k+q)\big[n(q)-n(k+q)\big]}
                             {c k \kappa^2/[2q(k+q)]+i\Gamma_{k12}}
\nonumber \\ &+&\int_o^kd q \frac{q(k-q)n(q)} {-c k
\kappa^2/[2q(k-q)]+i\Gamma_{k12}}
\Bigg] \ , 
\label{E81}
\end{eqnarray}
where
\begin{equation} 
A=3(g+1)\sqrt{c/4\pi^3\rho_0}\,,
\label {E82}
\end{equation}
is a factor in (\ref{A13}) so that for parallel or almost parallel
wavevectors $V({\bf k}, {\bf q},{\bf p}) =A\sqrt{k q p}$.  After
changing of variables this integral becomes to be more transparent:
\begin{eqnarray}
&&\!\!\!\!\!\!
\Sigma({\bf k})=\frac{A^2}{4\pi^2 c}
\Bigg[\int d q \int_0^{y_{\rm max}} d y 
        q^2(k+q)^2 
\label{E9}\\
&\times &\frac{[n(q)-n(k+q)]}{y+i\Gamma_{k12}}
-\int_0^{y_{\rm max}} 
        d y\int_o^k d q \frac{q^2(k-q)^2n(q)}{y-i\Gamma_{k12}}
\Bigg] \ .
\nonumber
\end{eqnarray}
One may estimate $y_{\rm max}\simeq c k^2 / 2 q $ from the fact that
our expressions were obtained by expanding in $ \kappa / k$, therefore
should be at least $\kappa<k$.

Now let us consider the imaginary and real part of $\Sigma$
separately. It is convenient to begin with $ \gamma({\bf k})=-{\rm
  Im}\Sigma({\bf k})$:
\begin{eqnarray}
 \gamma({\bf k})&\simeq& \frac{A^2}{4\pi^2 c}
\int_0^\infty  d y 
\Bigg[\int_0^\infty d q q^2(k+q)^2\Gamma_{k12}
\label{E91}\\
&\times&
\frac{
\big[n(q)-n(k+q)\big]} {y^2+\Gamma_{k12}^2}
+ \int_0^k d q \frac{q^2(k-q)^2n(q)\Gamma_{k12}}{y^2+\Gamma^2_{k12}}
\Bigg] \ . 
\nonumber
\end{eqnarray}
Here we changed the upper limit of integration: $y_{\rm max}\to\infty$
because the main contribution to the integral comes from the area
$y\simeq\Gamma\ll c k$.  After trivial integration with respect of $y$
one has:
\begin{eqnarray}
 \gamma({\bf k})&\simeq &  \frac{A^2}{8\pi c }\Bigg[ \int_0^\infty  
q^2(k+q)^2\big[n(q)-n(k+q)\big] d q 
\nonumber \\
&&+ \int_0^k  q^2(k-q)^2n(q) d q\Bigg] \ .
\label{E10}
\end{eqnarray}
This expression for $ \gamma({\bf k})$ corresponds to that given by
the kinetic equation \cite{ZLF} for waves.  For further progress it is
necessary to do some assumption about $n(q)$.  Let us assume that
$n(q)$ vanishes with growing of $q$ faster than
$1/q^4$.\footnote{\small Remember, that in Zakharov-Sagdeev spectrum
$n(q)\propto q^{-9/2}$, and in Kadomtsev-Petviashvili spectrum
$n(q)\propto q^{-4}$. This assumption is true for the Zakharov-Sagdeev
spectrum and is not true for Kadomtsev-Petviashvili one.} For such
spectra the main contribution to the integral comes from small $q\ll
k$.  In this case contributions from first and second integrals in
(\ref{E10}) coincides and may be represented in the form:
\begin{eqnarray} 
 \gamma(k)=\frac{A^2 k^2}{4 \pi c }\int_{1/L}^\infty
{n(q) q^2 d q}
\simeq 
\frac{A^2 k^2 }{4\pi c } N(\Omega) \ .
\label {E11aa}
\end{eqnarray}
In this case contributions from first and second integrals in
(\ref{E10}) coincides and may be represented in the form:
\begin{eqnarray} 
 \gamma(k)=\frac{A^2 k^2}{4 \pi c }\int_{1/L}^\infty
{n(q) q^2 d q}
\simeq 
\frac{A^2 k^2 }{4\pi c } N(\Omega) \ .
\label {E11a}
\end{eqnarray} 
\section{Estimation of the two-loop diagrams.}
Let us write down analytical expression which correspond to one of the
diagrams (b) in Fig.1(b)
\begin{eqnarray}
\Sigma_b({\bf k},\omega)&&= \int \frac{ d{\bf k_1} d {\bf k_2} 
d\omega_1 d\omega_2} {(2\pi)^8} V_a V_b V_c V_d
n(k_1,\omega_1)n(k_2,\omega_2)
\nonumber\\&&
\times G({\bf k_1}+{\bf k_2},
\omega_1+\omega_2)  G({\bf k}+{\bf k_1}+
{\bf k_2},\omega+\omega_1+\omega_2)
\nonumber\\
 &&\times G({\bf k}+{\bf k},\omega+\omega_2)\label{AppC1}
\end{eqnarray}
where $V_a, V_b, V_c, V_d$ are vertices,
\begin{eqnarray}
V_a=V({\bf k_1}+{\bf k_2}+{\bf k},{\bf k},{\bf k_1}+{\bf k_2}),\\
V_b=V({\bf k_1}+{\bf k_2}+{\bf k},{\bf k_1},{\bf k}+{\bf k_2}),\\
V_c=V({\bf k_2}+{\bf k},{\bf k_2},{\bf k}),\\ V_d=V({\bf k_1}+{\bf
k_2},{\bf k_1},{\bf k_2})\end{eqnarray} We just followed the rules of
DT and integrated over all delta-functions.  From now, the analyses
will be parallel to that of appendix B.  Let us use (\ref{D6}) for
$n(k,\omega)$ and (\ref{D2}) for $G(k,\omega)$.  Now we can easily
perform integration over $\omega_1$ and $\omega_2$.  Now, as it was
done in appendix B, introduce $\Sigma_b({\bf k})=\Sigma_b({\bf
k},\omega_*)$. Since all interacting wavevectors are almost parallel,
we introduce two-dimensional vectors $\bf \kappa_1$ and $\bf \kappa_2$
such that
\begin{eqnarray}
{\bf k}_1=q_1 {\bf k} /k +{\bf \kappa_1 }\,, \qquad {\bbox \kappa_1
}\perp{\bf k}\ \\ {\bf k}_2=q_2 {\bf k} /k +{\bf \kappa_2 }\,, \qquad
{\bbox \kappa_2 }\perp{\bf k}\ \\
\end{eqnarray}
We use $V({\bf k}, {\bf q},{\bf p}) =A\sqrt{k q p}$. Since
$\kappa_i\ll k$, we can expand resonance denominators in (\ref{AppC1})
with respect to $\kappa_i$. The integrals will be dominated by
regions, where $q_i\ll k$. By putting everything together, one gets
\begin{eqnarray}
\Sigma_b({\bf k})\simeq 
\int \frac{ \pi^2 d q_1 d q_2 d \kappa_1^2 d \kappa_2^2}{2\pi^6}
A^4 k^3 (q_1+q_2) q_1 q_2 \tilde n_{q_1} \tilde n_{q_2} \\
\left[ (\frac{c(\kappa_1^2+\kappa_2^2)}{2(q_1+q_2)}+
\Gamma_{q_1,q_2,q_1+q_2})
       (\frac{c}{2} \frac{\kappa_1^2}{q_1}+\frac{\kappa_2^2}{q_2}
+i\gamma_k)
\right. \nonumber \\ \left.
       (\frac{c \kappa_2^2}{2 q_2}+ i \gamma_k)\right]^{-1}
\end{eqnarray} Substituting $\tilde n_q=n/q^{-9/2}$ 
we see, that indeed, the dominant part comes from the region of small
$q_i$. We can estimate all these integrals to get
\begin{equation}\Sigma_b\simeq \frac{A^4 k^3 L^2 n^2}{c^2 \gamma_k}
\end{equation}
where we used the small $q$ cutoff $1/L$.  Finally
\begin{equation} 
\frac{\Sigma_b}{\gamma_k}\simeq \frac{k^3 L^2 n^2}{\rho_0^2 \gamma_k^2}
\simeq\frac{1}{k L}\ll 1, \end{equation}
and we conclude, that contribution from diagrams of type (b) on
Fig.1(b) is much less, than contribution from one-loop diagrams. But
this is not the end of the story.  Let us try to estimate
contributions from diagrams of type (e) on Fig.1(b).  Following the
same guidelines, we obtain
\begin{eqnarray}
\Sigma_e({\bf k})=&&
\int {\frac{d {\bf k_1} {\bf k_2}}{(2 \pi)^8}
 A^4 k k_1 k_2 (k+k_1) (k+k_1+k_2)(k+k_2)
\tilde n_{k_1} \tilde n_{k_2}} \nonumber\\ &&\times
\left[ 
(\omega_k+\omega_{k_1} -\omega_{k+k_1} + i\Gamma_{k,k_1,k+k_1})
\right.\nonumber \\ &&\left.\times
(\omega_k+\omega_{k_1}+\omega_{k_2} -\omega_{k+k_1+k_2} +
i\Gamma_{k,k_1,k_2,k+k_1+k_2})
\right.\nonumber \\ && \times\left.
(\omega_k+\omega_{k_2} -\omega_{k+k_2} +
i\Gamma_{k,k_1,k+k_2})\right]^{-1}
\nonumber
\end{eqnarray}
Let us again introduce $\kappa_1$ and $\kappa_2$ as above, and
substituting $\tilde n_q=n/q^{-9/2}$ we obtain the following
estimation:
\begin{equation} \frac{\Sigma_e}{\gamma_k}\simeq 
\frac{A^4 k^4 n^2 L^3}{\gamma_k^2 c^2 }\simeq 1 \end{equation}

Therefore we conclude, that contribution from two loop diagrams is
dominated by planar diagrams and of the order of one loop diagrams
contribution.  }

\end{multicols}
\end{document}